\documentclass[aps, 10pt, prd, longbibliography,
               notitlepage, twocolumn, superscriptaddress,
               nofootinbib, floatfix]{revtex4-2}

\usepackage[english]{babel}
\usepackage{amsmath}
\usepackage{amssymb}
\usepackage{amsfonts}
\usepackage[utf8]{inputenc}
\usepackage[T1]{fontenc}
\usepackage{mathrsfs}
\usepackage{anyfontsize}
\usepackage{microtype}
\usepackage{xspace}
\usepackage{multirow}
\usepackage{dsfont}
\usepackage{ltablex}
\usepackage{longtable}
\usepackage{placeins}

\usepackage[linktocpage,breaklinks]{hyperref}
\usepackage[dvipsnames]{xcolor}

\usepackage{tensor}
\usepackage{bm}

\usepackage{graphicx}
\usepackage{epsfig}
\usepackage{epstopdf}

\usepackage{natbib}
\usepackage{colortbl}
\usepackage{tabularx}
\usepackage{numprint}

\usepackage{CJKutf8}

\hypersetup{colorlinks=true,
            citecolor=NavyBlue,
            linkcolor=NavyBlue,
            urlcolor=NavyBlue}

\newcommand{\qand}{\quad\textrm{and}\quad}
\newcommand{\eos}{NL3$\omega\rho$L55~}

\newcommand{\msun}{\textrm{M}_{\odot}}
\newcommand{\mc}{m_{\chi}}
\newcommand{\mb}{m_{\rm b}}
\newcommand{\mdm}{m_{\rm d}}

\newcommand{\pb}{p_{\rm b}}

\newcommand{\pdm}{p_{\rm d}}
\newcommand{\dd}{{\rm d}}
\newcommand{\ee}{{\rm e}}
\newcommand{\ii}{{\rm i}}
\newcommand{\bb}{{\rm b}}
\newcommand{\cc}{{\rm c}}
\newcommand{\hh}{{\rm h}}
\newcommand{\su}{{\rm s}}
\newcommand{\tl}{{\rm t}}
\newcommand{\dm}{\delta_{\rm M}}

\newcommand{\tfdm}{{}_{2}t_{0}}

\def\newacronym#1#2#3{\gdef#1{#2\xspace}}
\newacronym{\DANS}{DANS}{dark-matter-admixed neutron star}
\newacronym{\DM}{DM}{dark matter}
\newacronym{\PNS}{PNS}{pure neutron star}

\newcommand{\uiuc}{\affiliation{Department of Physics and
Illinois Center for Advanced Studies of the Universe,\\
The Grainger College of Engineering,
University of Illinois Urbana-Champaign, Urbana, Illinois 61801, USA}}

\begin{document}
\title{The crust of dark-matter admixed neutron stars:\\
bulk properties and torsional oscillations}

\begin{abstract}
We study how dark matter impacts the crust and the spectrum of torsional crust
oscillations of dark-matter-admixed neutron stars.
We construct two-fluid equilibrium solutions wherein baryonic and dark matter
interact gravitationally only, adopting a unified nuclear equation of state for
the former and a fermionic equation of state with repulsive self-interaction
for the latter.
At fixed total gravitational mass and dark-matter mass fraction, we find that
dark matter reduces the crust thickness in comparison to pure baryonic-matter
neutron stars.
The thinning of the crust is negligible when most of the dark matter
distribution extends beyond the star's baryonic surface.
However, the crust thickness can decrease by as much as 12\% when the dark matter
distribution is within the star's baryonic surface, i.e., when the star has a ``dark core.''
We support these results by deriving approximate analytical formulas for
the crust thickness that agree with our numerical calculations at the sub-percent level in best case scenarios.
Next, we derive the equation that describes crustal torsional modes of
dark-matter-admixed neutron stars in the relativistic Cowling approximation.
We find that the oscillation frequencies are in general higher than
those of a comparable pure baryonic-matter neutron star, with the largest frequency shifts 
happening in the same parameter space where the crust thickness decreases the most.
Moreover, we study the degeneracy between dark-matter and baryonic-crustal
microphysics effects on these modes. As an example, we study electron screening,
which softens the crust's shear modulus, thus decreasing the frequencies.
We find that the degeneracy between the competing effects of dark matter and electron screening 
can be broken in some regions of the parameter space we explored, e.g., for massive 
$2\,\msun$ neutron stars with compact dark-matter distributions.
Should they be measured, our results suggest that torsional oscillations
could be used to infer the existence of a dark matter core within massive neutron
stars.
\end{abstract}

\author{Jiayi Zhang\begin{CJK*}{UTF8}{gbsn}（张嘉懿）\end{CJK*}} \email{jz124@illinois.edu}   \uiuc
\author{Hector O. Silva}                                       \email{hosilva@illinois.edu} \uiuc
\maketitle

\section{Introduction} \label{sec:intro}

Neutron stars provide a unique laboratory for studying matter under conditions inaccessible to terrestrial experiments~\cite{Lattimer:2004pg,Ozel:2016oaf,Baym:2017whm,Lattimer:2021emm}. 
Their large compactness, high baryon density, and long lifetimes also make them natural environments in which dark matter may accumulate through scattering and gravitational capture over astrophysical timescales~\cite{Goldman:1989nd,Bell:2020jou,Busoni:2021zoe}.
If the captured dark matter is stable or non-annihilating, it may accumulate inside or around the star, motivating the study of dark-matter-admixed neutron stars (\DANS{s})~\cite{Sandin:2008db,Ciarcelluti:2010ji,Rutherford:2022xeb,Grippa:2024ach,Rutherford:2024uix}. 
Because dark matter can affect the star's structure~\cite{Leung:2011zz,Das:2020ecp,Leung:2022wcf,Kumar:2024zzl,Kumar:2025ytm}, 
it is crucial to identify observables that could probe its presence in neutron stars.

Neutron star asteroseismology offers one such probe.
In particular, quasiperiodic oscillations (QPOs) observed in the decaying tails of giant magnetar flares have been discussed in connection with torsional oscillations of the solid neutron star crust and, more generally, global magneto-elastic modes; see, e.g., Refs.~\cite{Duncan:1998my,Israel:2005av,Strohmayer:2006py,Watts:2006mr,Levin:2006qd,El-Mezeini:2010xxh}. 
Since the properties of the torsional modes are controlled by properties of the crust, such as its elasticity and thickness~\cite{Schumaker:1983MNRAS.203..457S,Samuelsson:2006tt,Kozhberov:2020lzm,Yakovlev:2022nui}, these modes provide a direct link between microphysics and observations;
see Refs.~\cite{Watts:2011kh,Sotani:2024mlb} for reviews.
For \DANS{}, this naturally raises the question: does dark matter leave seismic imprints 
on the crustal torsional oscillation modes?
 
We give an initial positive answer to this question by studying the properties of the (baryonic) crust and the torsional oscillation spectrum of \DANS{s}, constructed in the two-fluid formalism~\cite{Kodama:PTP1972,Sandin:2008db,Ciarcelluti:2010ji}.
We study how dark matter modifies the crust thickness and how these changes are reflected in the fundamental and first-overtone torsional oscillation frequencies. We  also compare the frequency shifts induced by dark matter with those due to electron screening in the crustal shear modulus~\cite{Kobyakov:2013eta}. This represents one aspect of crustal baryonic microphysics that is known to also impact the frequencies of these modes~\cite{Sotani:2014dua}.

This paper is organized as follows. In Sec.~\ref{sec:dans}, we introduce the two-fluid formalism
we use to model \DANS{s}, present the baryonic and dark matter equations of state we adopt, and describe how we numerically construct \DANS{} solutions.
In Sec.~\ref{sec:crust}, we study the structure of the crust of DANS{s} and present approximate analytical formulas for the crust thickness.
In Sec.~\ref{sec:tors}, we derive the master equation governing torsional oscillations in the relativistic Cowling approximation and describe the elastic properties of the crust.
In Sec.~\ref{sec:results}, we study the torsional mode spectrum for \DANS{s} and compare it with the spectrum of pure baryonic matter neutron stars (\PNS{s}). 
In Sec.~\ref{sec:results_es}, we include electron screening in the shear modulus, and discuss the degeneracy between crust-microphysics and dark-matter-induced frequency shifts.
We summarize our results, discuss their limitations, and outline directions for further studies 
in Sec.~\ref{sec:discussions}.
Three appendices complement the main text. In Appendix~\ref{app:crust_thickness}, we present the derivation of 
the analytical formulas for crust thickness, and assess their accuracy in Appendix~\ref{app:ct_table}.
At last, we tabulate all oscillation frequencies computed herein in Appendix~\ref{app:tor_num_results}.

We work in units in which $c = G = \hbar = 1$ and adopt the mostly 
plus metric signature.

\section{Dark matter admixed neutron stars} \label{sec:dans}

\subsection{Equations of stellar structure in the two-fluid formalism} \label{sec:dans:tov}

We consider a static and spherically symmetric star, whose spacetime
can be described by the line element:
\begin{equation} \label{eq:line_element}
    \dd s^2 =
    - \ee^{2\Phi}    \, \dd t^2
    + \ee^{2\Lambda} \, \dd r^2
    + r^2 (\dd \theta^2
    + \sin^2\theta \, \dd \phi^2),
\end{equation}
where $\Phi$ and $\Lambda$ are functions of $r$. 
The star is modeled as a system of two perfect fluids, describing baryonic and dark matter, which we assume interact only gravitationally.
The energy-momentum tensor of a perfect fluid is
\begin{equation}\label{eq:T_perfect_fluid}
    T_{\mu\nu} = (\varepsilon+p) u_\mu u_\nu + p g_{\mu\nu},
\end{equation}
where $\varepsilon$ is the total energy density, $p$ is the pressure,
$u^{\mu}$ is the fluid's four-velocity, satisfying $u_{\alpha} u^{\alpha} = -1$, and $g_{\mu\nu}$ is the spacetime
metric which can be read off Eq.~\eqref{eq:line_element}.

Throughout this work, we use the subscripts ``b'' and ``d'' to indicate a quantity
related to baryonic and dark matter respectively.
We use the subscript ``t'' (for total) to indicate the sum of the same quantity
for the baryonic and dark-matter fluid components.
For example, the total energy density is $\varepsilon_\tl = \varepsilon_\bb + \varepsilon_\dd$.

References~\cite{Sandin:2008db,Ciarcelluti:2010ji} (see also the earlier work~\cite{Kodama:PTP1972}) showed that a two-component
fluid star is described by a Tolman--Oppenheimer--Volkoff system of equations~\cite{Tolman:1939jz,Oppenheimer:1939ne}, namely,
\begingroup
\allowdisplaybreaks
\begin{subequations} \label{eq:tov_eqs_all}
    \begin{align}
        \frac{\dd \Phi}{\dd r} &=\frac{m_\tl +4\pi r^3 p_\tl}{r (r - 2 m_\tl )},
        \label{eq:tov_phi}
        \\
        \frac{\dd \pb}{\dd r}  &= - (\varepsilon_\bb + \pb) \frac{\dd \Phi}{\dd r},
        \label{eq:tov_pb}
        \\
        \frac{\dd \pdm}{\dd r} &= - (\varepsilon_\dd + \pdm) \frac{\dd \Phi}{\dd r},
        \\
        \frac{\dd \mb}{\dd r}  &= 4 \pi r^2 \varepsilon_{\bb},
        \\
        \frac{\dd \mdm}{\dd r} &= 4 \pi r^2 \varepsilon_{\dd}.
    \end{align}
\end{subequations}
\endgroup
Here,  $p_\tl = p_\bb + p_\dd$ is the total pressure and
$m_\tl = m_\bb + m_\dd$ is the total mass function inside a sphere of radius $r$,
related to the metric function $\Lambda$ as
\begin{equation}
\Lambda = - \tfrac{1}{2} \ln(1 - 2m_\tl / r).
\end{equation}

The system of equations~\eqref{eq:tov_eqs_all} has five equations for seven variables.
To close the system, we need two equations of state that relate
the energy density and pressure of each fluid. We describe them next.

\subsection{Equations of state} \label{sec:dans:eos}

\subsubsection{The baryonic equation of state} \label{sec:dans:eos:beos}

We adopt the \eos equation of state~\cite{Horowitz:2000xj,Pais:2016xiu},
obtained from the CompOSE database~\cite{Typel:2013rza,Oertel:2016bki,CompOSECoreTeam:2022ddl}.
The data tables we use in our numerical calculations are generated as described 
in Shawqi and Morsink~\cite{Shawqi:2024jmk}, Sec.~2.1.

\eos is a stiff equation of state that predicts a maximum mass $M_{\rm max} =
2.752\,\msun$ for a nonrotating neutron star.
It is also a unified equation of state, in the sense that the outer and inner
crusts and liquid core are all calculated within the same underlying
nuclear interaction model instead of being matched piecewise from
different models~\cite{Potekhin:2013qqa}.
We use an unified equation of state because core-crust matching done with
nonunified equations of state can introduce large systematic uncertainties in
the neutron-star structure, that can be as larger as $30\%$ in the crust
thickness and $4\%$ in the radius~\cite{Fortin:2016hny}.
According to Ref.~\cite{Pais:2016xiu}, Table~II, the crust-core interface
happens at baryon number density $n_{{\rm b},{\rm cc}} = 0.084\,\textrm{fm}^{-3}$,
which corresponds to the pressure
\begin{equation} \label{eq:crust_core_pres}
    p_{{\rm b},{\rm cc}} = 0.516\,\mathrm{MeV\,fm^{-3}}
    \quad (8.27\times10^{32}\,\mathrm{dyn\,cm^{-2}}).
\end{equation}

\subsubsection{The dark matter equation of state} \label{sec:dans:eos:dmeos}

As an illustrative example, we adopt a fermionic dark-matter equation of state 
with repulsive self-interaction~\cite{Narain:2006kx,Nelson:2018xtr}.
This equation of state was employed in several works in the literature,
including, e.g., Refs.~\cite{Miao:2022rqj,Routaray:2024fcq,Konstantinou:2024ynd,Shawqi:2025cca,Zhou:2025dmy}.

The energy density, $\varepsilon_{\dd}$, and pressure, $\pdm$, are:
\begingroup
\allowdisplaybreaks
\begin{subequations} \label{eq:eos_dm}
    \begin{align}
        \varepsilon_{\dd} &= \frac{\mc^4}{8 \pi^2}
        \, [(2 x^3+x) \sqrt{1+x^2} - \textrm{arcsinh}\,x]
        \nonumber \\
        \, &\quad + \frac{\mc^4 \, y^2}{(3\pi^2)^2} \, x^6,
        \label{eq:eos_dm_e}
        \\
        \pdm &= \frac{\mc^4}{24 \pi^2}
        \, [(2 x^3 - 3x) \sqrt{1+x^2} + 3 \, \textrm{arcsinh}\,x]
        \nonumber \\
        \, &\quad
        \, + \frac{\mc^4 \, y^2}{(3\pi^2)^2} \, x^6,
        \label{eq:eos_dm_p}
    \end{align}
\end{subequations}
\endgroup
where $\mc$ is the dark-matter particle mass;
$x = k_\mathrm{F}/\mc$ is the dimensionless ``relativistic parameter'' defined in terms of the Fermi momentum $ k_\mathrm{F}$;
and $y=\mc/m_I$ is a dimensionless quantity that characterizes
the repulsive self-interaction strength, where $m_I$ is the mass of the self-interaction force mediator.
Following Ref.~\cite{Shawqi:2024jmk}, we consider dark-matter particle masses and
self-interaction strengths in the ranges,
\begin{equation} \label{eq:dm_param_range}
    m_\chi \in [50,\,1000]\,\textrm{MeV},
    \qand
    y \in [0,\,80].
\end{equation}

\subsection{Numerical integration} \label{sec:dans:numerics}

To obtain two-fluid star solutions we integrate the coupled system of differential equations~\eqref{eq:tov_eqs_all} outward, starting from the center of the star, using the Runge--Kutta--Fehlberg method.
Each solution can be parametrized by its central baryonic and dark-matter energy densities, $\varepsilon_{\bb,\cc}$ and $\varepsilon_{\dd,\cc}$. The latter can be determined from Eq.~\eqref{eq:eos_dm_e}, for given values of $m_\chi$, $y$, and $x_\cc$.

The integration is carried out until both fluids reach their respective surfaces,
located at radii $R_\bb$ and $R_\dd$.
We again follow Ref.~\cite{Shawqi:2024jmk}, and locate these radii through the conditions,
\begin{equation}
    (\dd m_\bb / \dd r)|_{r=R_\bb} = 0,
    \qand
    (\dd m_\dd / \dd r)|_{r=R_\dd} = 0.
\end{equation}
(Alternatively, one could also identify $R_\bb$ and $R_\dd$ by
finding the locations where the pressures $p_\bb$ and $p_\dd$ vanish.)
The mass functions evaluated at $R_\bb$ and $R_\dd$ give the total
baryonic and dark-matter masses,
\begin{equation}
    M_\bb = m_\bb(R_\bb),
    \qand
    M_\dd = m_\dd(R_\dd),
\end{equation}
and their sum defines the total gravitational mass of the system,
\begin{equation}
    M_\tl = M_\bb + M_\dd.
\end{equation}

If $R_\dd$ is smaller than $R_\bb$, the star has a ``dark core.'' In this case,
we continue the integration for $r > R_\dd$, setting  $p_\dd = \varepsilon_\dd = 0$,
until we reach $R_\bb$.
Conversely, if $R_\bb$ is smaller than $R_\dd$, the star has a ``dark matter halo.''
In this case, we continue the integration for $r > R_\bb$, setting
$p_\bb = \varepsilon_\bb = 0$, until we reach $R_\dd$.

While doing the integration, we also monitor where the baryonic pressure
reaches the crust-core interface value~\eqref{eq:crust_core_pres}.
This defines the location of the crust-core interface, $R_{\rm cc}$,
\begin{equation}
    p_\bb(R_{\rm cc}) = p_{{\rm b},{\rm cc}}.
\end{equation}
The dark-matter fluid component has no solid crust.
For this reason, we omit a subscript ``b'' from $R_{\rm cc}$.

In our study, we focus on configurations with fixed total mass $M_\tl$ and dark-matter mass fraction,
\begin{equation} \label{eq:def_dm_mass_frac}
    f_\chi = M_\dd/M_\tl.
\end{equation}
Throughout this work we use
\begin{equation}
    f_{\chi} = 0.05,
\end{equation}
to explore the implications of a relatively large dark matter fraction,
as suggested by the \DANS formation
scenarios discussed in Ref.~\cite{Ellis:2018bkr}.

For a chosen value of $M_\tl$, we can then perform a two-dimensional root-finding
search to determine the pair $(\varepsilon_{\bb,\cc},\,\varepsilon_{\dd,\cc})$
that yields a star with the desired values of $M_{\tl}$ and $f_{\chi}$.
We validated our numerical code by reproducing the stellar-structure results of Ref.~\cite{Shawqi:2024jmk}.

\subsection{Solution space} \label{sec:dans:sols}

\begin{table}[t]
\caption{Types of \DANS{} configurations classified according
to their halo-mass fraction, $\delta_{\rm M}$, defined as the ratio
between the gravitational dark-matter halo and total masses, $M_\hh$ and $M_\tl$.}
\begin{tabular}{ l c c r}
\arrayrulecolor{Gray}
\hline \hline
Stellar configuration & & & $\delta_{\rm M}$  \\
\hline
Dark core         & & & 0 \\
Compact halo      & & & $0 < \dm < 0.6$ \\
Intermediate halo & & & $0.6 \leq \dm < 0.85$ \\
Diffuse halo      & & & $0.85 \leq \dm < 1$ \\
\hline \hline
\end{tabular}
\label{tab:dans_class}
\end{table}

For each solution, we also calculate other quantities of interest.
We define the dark-matter halo mass, $M_{\hh}$, as the difference between the
total mass, $M_\tl$, and the total mass function evaluated at the baryonic radius $R_{\bb}$, i.e.,
\begin{equation}
    M_{\hh} = M_\tl - m_\tl(R_{\bb}).
\end{equation}
To characterize the \DANS{} solution space, it is convenient
to introduce the halo-mass fraction,
\begin{equation}
    \dm = M_{\hh} / M_{\dd}.
\end{equation}
If this quantity is zero, all dark matter is confined within the baryonic
radius of the neutron star --- we say the star has a dark core.
If $\dm$ is nonzero, the neutron star has a dark matter halo.
We follow Ref.~\cite{Shawqi:2024jmk} that suggested three types of halo configurations
according to the value of $\dm$.
We say the star has a ``compact halo'' if $0 < \dm < 0.6$.
If $0.6 \leq \dm < 0.85$ we say the star has an ``intermediate halo,''
and, finally, if $0.85 \leq \dm < 1$ we say the star has a ``diffuse halo.''
We summarize this classification in Table~\ref{tab:dans_class}.

To explore the solution space, we calculated the mass-radius
relation for several values of $\mc$, but neglecting self-interaction, $y$.
The reason why we neglect it will be explained
in Sec.~\ref{sec:dans:crust} when we discuss the crust thickness
of \DANS{s}.

\begin{figure}[t]
    \includegraphics[width=\columnwidth]{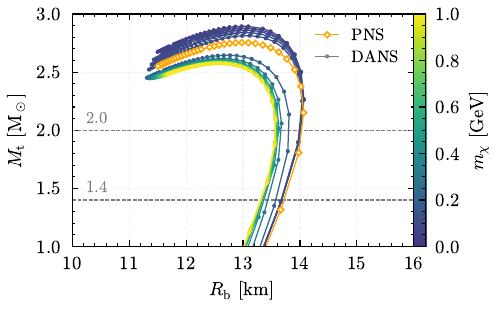}
    \caption{
    Bulk properties of \DANS{s}. 
    We show the total mass, $M_\tl$, as a function of the baryonic radius, $R_\bb$, for 
    stars
    with fixed dark-matter mass fraction $f_\chi = 0.05$. The color gradient across the different curves with circle markers indicates the 
    value of the dark matter mass $\mc$.
    The curve with diamond markers shows 
    the mass-radius relation of \PNS{s} modeled with the \eos equation of state.
    The horizontal dashed lines mark the two total masses, $M_\tl=1.4\,\msun$ and $M_\tl=2.0\,\msun$.
    We see that the larger $\mc$ becomes, the larger are the deviations away from the \PNS{} sequence.
    However, a clustering of curves is noticeable as $\mc$ approaches $1$~GeV.}
    \label{fig:mass_radius}
\end{figure}

In Fig.~\ref{fig:mass_radius}, we show the total mass $M_\tl$ as a
function of the baryonic radius $R_\bb$ for different values of
$\mc$ in the range~\eqref{eq:dm_param_range}.
Along each sequence of solutions, we vary the central value of the
baryonic energy density in the range
\begin{equation}
\varepsilon_{\bb,\cc} \in [0.2,\,4.0]\times10^{15} \, \textrm{g/cm}^{3},
\end{equation}
in increments of $\Delta \varepsilon_{\bb,\cc} = 0.1\times10^{15}$\,g/cm$^{3}$,
and search for the value of $\varepsilon_{\dd,\cc}$ that yields
a solution with dark-matter mass fraction $f_\chi = 0.05$.
Each mass-radius curve is colored according to the value of $m_\chi$, as indicated in the colorbar.
For comparison, we also show the mass-radius relation for a
pure baryonic matter neutron star (\PNS{}).
In general, heavier
dark matter particle masses result
in larger deviations in the mass-radius curves relative to the \PNS{} one.
However, this trend is saturated as $\mc$ approaches 1~GeV.
Finally, the two horizontal lines indicate 
total masses
$M_\tl = 1.4\,\msun$ and $2\,\msun$; these will be the
total masses of the \DANS{} solutions whose
torsional oscillation frequencies we will investigate in
Sec.~\ref{sec:results}.

\begin{figure}[t]
    \includegraphics[width=0.965\columnwidth]{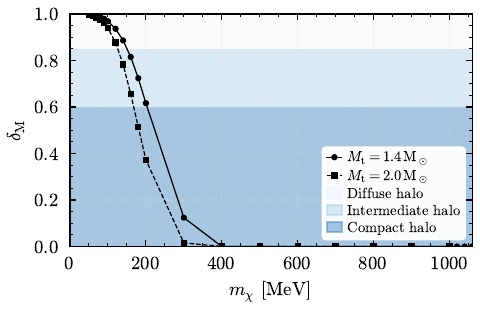}
    \caption{The halo-mass fraction $\delta_{\rm M}$ as a function of $\mc$ for the two fixed-mass \DANS{} sequences. The shaded regions indicate the diffuse, intermediate, and compact halo regimes, classified in Table~\ref{tab:dans_class}; the dark core region $\delta_{\rm M} = 0$ is not visible above. The solid and dashed curves represent the $M_\tl=1.4\,\msun$ and $M_\tl=2.0\,\msun$ sequences, respectively. We see that as $\mc$ increases, the solutions transit from diffuse-halo to dark-core configurations.
    The shading pattern we use here to indicate different \DANS{} configurations will be used in most figures in the remainder of this work.}
    \label{fig:halo_fraction}
\end{figure}

Complementary to the previous discussion, in Fig~\ref{fig:halo_fraction} we
show the halo-mass fraction $\delta_{\rm M}$ for the two sequences of solutions of
fixed total masses, $M_\tl = 1.4\,\msun$ (solid line) and $2\,\msun$ (dashed line), as a function of the dark-matter particle mass $\mc$.
To guide the eye, we shade with different colors the intervals of $\delta_{\rm M}$ corresponding
to the \DANS{} configurations we classified in Table~\ref{tab:dans_class}; this will be a recurring pattern in most figures in our work.
We see that for both curves, the \DANS{} harbors
a diffuse halo when $\mc \lesssim 160$\,MeV,
intermediate halo when $160 \lesssim \mc \lesssim 200$\,MeV,
compact halo when $200 \lesssim \mc \lesssim 400$\,MeV,
and contains a dark core for $\mc \gtrsim 400$\,MeV.
These results are important for the remainder of our work. As we will show in Sec.~\ref{sec:crust},
they correlate with how dark matter impacts the crust of \DANS{} solutions 
relative to their \PNS cousins.

Since the solutions in Fig.~\ref{fig:halo_fraction} are the ones we will
study hereafter, let us briefly comment on their stability.
As discussed in Ref.~\cite{Shawqi:2024jmk} (based on
Refs.~\cite{Sorkin:1982ut,Jetzer:1990xa,Henriques:1990xg}; see also Refs.~\cite{Kain:2021hpk,Gleason:2022eeg,Caballero:2024qtv}),
a \DANS{} is stable to linear radial perturbations if it is situated at or to the right of the maximum mass on
a $M_\tl$ versus $R_\bb$ curve of constant $f_\chi$.
Figure~\ref{fig:mass_radius} shows that all solutions we will study are stable.

\section{The crust of dark-matter admixed neutron stars} \label{sec:crust}

\subsection{General properties of the crust} \label{sec:dans:crust}

The crust thickness,
\begin{equation}
    \Delta R = R_{\bb} - R_{{\rm cc}},
\end{equation}
is important in determining the torsional oscillation frequencies of single
fluid neutron stars~\cite{Samuelsson:2006tt}, and in various other phenomena
involving neutron stars~\cite{Chamel:2008ca}.

\begin{figure}[t]
    \includegraphics[width=\columnwidth]{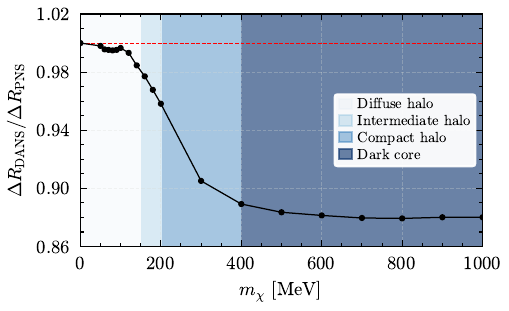}
    \caption{The crust thickness of \DANS{s}
    with $M_\tl = 1.4\,\msun$, normalized with respect to the \PNS{} value,
    $\Delta R_{\rm PNS} = 1.556$~km,
    as a function of $m_\chi$ and vanishing dark matter self-interaction. The horizontal dashed line corresponds
    to the \PNS{} value.
    The shaded regions indicate the different \DANS{s}
    configurations, from diffuse halo to dark core. We see that for small values of 
    $\mc$, the crust thickness of \DANS{s} changes little with respect to the \PNS{} case.
    The crust becomes progressively thinner as $\mc$ increases, saturating as $\mc$ approaches 1~GeV.}
    \label{fig:crust_ratio}
\end{figure}

How does the presence of dark matter affects the crust thickness?
To answer this question, we continue neglecting self-interaction in the dark matter equation of state, and investigate how the crust thickness $\Delta R$ depends on $\mc$.
We show our findings in Fig.~\ref{fig:crust_ratio} for
the sequence of solutions with $M_\tl = 1.4\,\msun$, where we normalized the \DANS{} crust thickness $\Delta R_{\rm DANS}$ by that of a \PNS{} of mass $1.33\,\msun$, namely, $\Delta R_{\rm PNS} = 1.556~\mathrm{km}$.
Why do we compare our \DANS{} solutions to a \PNS{} of different mass?
The reason is that the baryonic fluid component contributes
\begin{equation} \label{eq:mb_for_dans}
    M_\bb = (1-f_\chi) \, M_\tl,
\end{equation}
to the total mass of the \DANS{}. Hence, a \PNS{} solution with a gravitational mass given by Eq.~\eqref{eq:mb_for_dans} is the correct reference star to compare against our \DANS{} solutions
in the limit in which $m_\chi$ vanishes.
Because we fix $f_\chi = 0.05$, we find that our reference \PNS{} solutions have
gravitational masses of $1.33\,\msun$ (to be compared against \DANS{} solutions with $M_\tl=1.4\,\msun$)
and $M_\tl=1.9\,\msun$ (to be compared against \DANS{} solutions with $M_\tl=2\,\msun$).

Resuming our discussion of Fig.~\ref{fig:crust_ratio}, we see
that for small $\mc$, the normalized crust thickness ratio tends toward $1$,
as expected.
As $\mc$ increases, the crust thickness of the \DANS{} solutions
become progressively thinner, with the largest gradient happening
as we cross the compact halo region, $200 \lesssim \mc \lesssim 400$\,MeV.
When we approach the dark core region, $\mc \gtrsim 400$~MeV,
the crust thickness becomes approximately constant, and
asymptotes to an approximate 12\% crust-thickness reduction
when $\mc$ reaches 1\,GeV.
We found qualitatively similar results for \DANS{} solutions
with $M_\tl=2.0\,\msun$. The largest crust thinning in this case becomes
approximately 16\% and happens, again, as $\mc$ approaches 1\,GeV.

\begin{figure}[t]
    \includegraphics[width=\columnwidth]{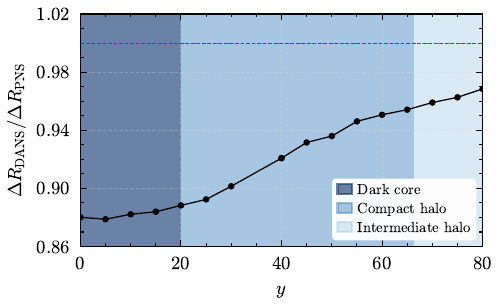}
    \caption{
    The crust thickness of \DANS{s} with $M_\tl =1.4\,\msun$ and $\mc = 1$~GeV normalized with respect to the \PNS{} value, $\Delta R_{\rm PNS} = 1.556$~km, as a function of the repulsive self-interaction strength $y$. As in Fig.~\ref{fig:crust_ratio}, the shaded regions indicate different \DANS{} configurations. 
    As $y$ increases, the solutions transition from dark-core to halo configurations, and the curve approaches 
    $1$, indicated with the dashed line. That is, as the self-interaction strength becomes stronger, the effects 
    of dark matter on the crust thickness are reduced.}
    \label{fig:crust_ratio_y}
\end{figure}

So far, we have only studied non-self-interacting dark matter.
What happens when $y$ becomes nonzero? In Fig.~\ref{fig:crust_ratio_y},
we again show the normalized crust thickness $\Delta R_{\rm DANS}$ with $M_\tl = 1.4\,\msun$,
but now for $\mc = 1$\,GeV and as a function of $y$ in the range~\eqref{eq:dm_param_range}.
We choose $\mc = 1$\,GeV because it is in this case that the deviations from
the \PNS{} case are largest, cf. Fig.~\ref{fig:crust_ratio}.
Figure~\ref{fig:crust_ratio_y} shows that as $y$ increases, the crust thickness
tends toward the \PNS{} value.
This can be understood as follows. Because self-interaction is repulsive,
the stronger it gets the less centrally concentrated the dark matter distribution
becomes. This ``puffs out'' the dark matter distribution in the \DANS{},
causing it to transition from a dark-core to a halo configuration.
In Fig.~\ref{fig:crust_ratio}, we saw that the crust thickness deviates by approximately 
5\% for intermediate-halo \DANS{s} and by even less for diffuse halo configurations. Thus,
by increasing $y$, we make the crust thickness approach that of our reference \PNS{} solution.
These observations are shared by the \DANS{} sequence of solutions with $M_\tl = 2\,\msun$.

The lessons of this subsection are clear. To maximize the deviations in the
crust thickness that, as we will see, maximize the deviations in the torsional
oscillation frequencies, we must have dark-core \DANS{s}.
Diffuse and intermediate dark-matter halos have small impact in the crust
thickness. Hence, while we will still explore the full range of values of $\mc$
for the sake of completeness, \emph{we will set the self-interaction strength, $y$,
to zero}, hereafter.

\subsection{Approximate analytical formulas for the crust thickness}\label{sec:ct_estimate}

In this section, we present approximate analytical formulas for the crust thickness of
\DANS{}. These formulas are inspired by existing ones for neutron stars; see,
in particular, Refs.~\cite{Samuelsson:2006tt,Baym:2017whm}.

Our main result is the following: the fractional crust thickness of a \DANS can
be approximated by:
\begin{align} \label{eq:ct_b6}
    \frac{\Delta R}{R_\bb}
    \approx
    \frac{
    \alpha \, (1-2C_\bb)
    }{
    C_\bb + \alpha \, (1-2C_\bb)
    },
    \,\,\,
    \textrm{with}
    \,\,\,
    \alpha = \ln(\mu_{\rm cc} / \mu_\su).
\end{align}
Here, $\mu_{\textrm{cc}}$ and $\mu_\su$ are the baryonic chemical potentials at the crust-core interface, $R_{\rm cc}$, and the baryonic surface, $R_\bb$, respectively.
For the \eos{}equation of state, $\alpha=0.0256$.
We also introduced the baryonic compactness,
\begin{equation} \label{eq:def_compactness}
    C_\bb = \frac{m_\tl(R_\bb)}{R_\bb}
          = \frac{M_\tl}{R_\bb} \left[ 1 - f_\chi + \frac{m_\dd(R_\bb)}{M_{\tl}} \right],
\end{equation}
where the second equality follows from using Eq.~\eqref{eq:mb_for_dans}.
We use the adjective ``baryonic,'' because $C_\bb$ depends on the \emph{total mass function $m_\tl$ evaluated
at $R_\bb$.}
A detailed derivation of Eq.~\eqref{eq:ct_b6}, together with a comparison to other
crust-thickness formulas, is presented in Appendix~\ref{app:crust_thickness}.

\begin{figure}[t]
    \includegraphics[width=\columnwidth]{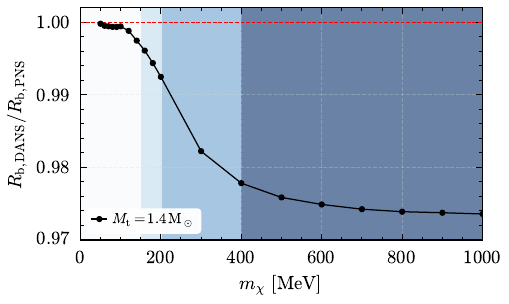}
    \caption{The baryonic radius $R_\bb$ of \DANS{s} normalized with respect to the corresponding \PNS{} value $R_{\bb,{\rm PNS}} = 13.702$~km as a function of dark matter particle mass $\mc$ for solutions with fixed total
    mass $M_\tl =1.4\,\msun$ and dark matter mass fraction $f_\chi=0.05$. 
    The shaded regions indicate different \DANS{} configurations, classified by the halo-mass fraction $\delta _M$. For small values of $\mc$ the baryonic radii of the \DANS{s} are nearly indistinguishable from that of the \PNS{}. As $\mc$ increases, $R_{\bb,{\rm DANS}}$ becomes progressively smaller than $R_{\bb,{\rm PNS}}$.}
    \label{fig:rb_pns_ratio}
\end{figure}

To build intuition on the behavior of $\Delta R / R_\bb$, we must first understand 
how $R_\bb$ and $C_\bb$ behave for \DANS{} solutions. We begin with the former.
In Fig.~\ref{fig:rb_pns_ratio}, we show the baryonic radius of our \DANS{} solutions with $M_\tl =1.4\,\msun$, normalized with respect to the radius of the reference \PNS, $R_{\bb} = 13.702$~km, as a function of  $\mc$. 
We see that $R_\bb$ decreases as $\mc$ increases, although with a smaller rate as $\mc$ approaches 1~GeV. This is consistent with the clustering of mass-radius curves we saw in Fig.~\ref{fig:mass_radius}. We also observed the same qualitative trend for the $M_\tl=2\,\msun$ sequence of solutions.

We now turn to the baryonic compactness.
Equation~\eqref{eq:def_compactness} takes  simple forms in the two extreme \DANS{} configurations: the dark-core and diffuse-halo limits. In the dark core limit, the dark-matter fluid is fully enclosed by the baryonic radius, so that
\begin{equation}
    m_\dd(R_\bb)=M_\dd=f_\chi \, M_\tl.
\end{equation}
The baryonic compactness~\eqref{eq:def_compactness} then reduces to
\begin{equation}\label{eq:cb_dc}
    C_{\bb,\mathrm{dc}} = {M_\tl} / {R_\bb};
\end{equation}
that is, $C_{\bb}$ is total gravitational mass enclosed within $R_\bb$.
In the diffuse halo regime, only a negligible fraction of the dark matter fluid
lies inside the baryonic radius; see, e.g., Ref.~\cite{Shawqi:2024jmk}, Fig.~1.
Hence,
\begin{equation}
    m_\dd(R_\bb) \ll M_\dd,
\end{equation}
and then,
\begin{equation} \label{eq:cb_dh}
    C_{\bb,\mathrm{dh}} \approx {(1-f_\chi)M_\tl} / {R_\bb},
\end{equation}
which is smaller than $C_{\bb,\mathrm{dc}}$ 
by an amount  $M_\dd / R_\bb$.

\begin{figure}[t]
    \includegraphics[width=0.972\columnwidth]{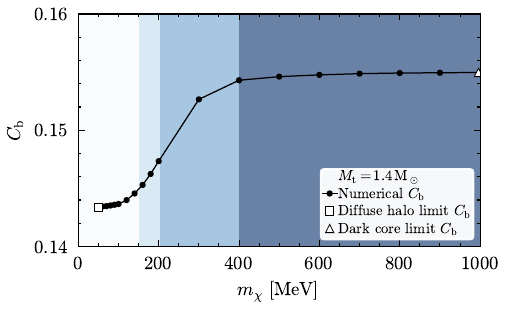}
    \caption{The baryonic compactness $C_\bb$ as a function of the dark matter particle mass $\mc$ for \DANS{} models with fixed total mass $M_\tl=1.4\,\msun$ and dark-matter mass fraction $f_\chi=0.05$.
    The curve shows the numerical results, while the square and triangle denote the diffuse-halo~\eqref{eq:cb_dh} and dark-core~\eqref{eq:cb_dc} limiting estimates, respectively. Both estimates agree well with the numerical data. As in previous figures, the shaded regions indicate the different \DANS{} configurations, classified by the halo-mass fraction $\delta_M$.}
    \label{fig:Cb}
\end{figure}

This conclusion is evident
when we plot $C_\bb$ as a function of $m_\chi$,
as we do in Fig.~\ref{fig:Cb}. 
For small values of $m_\chi$ (i.e., diffuse halo regime) the numerical values of $C_\bb$ remain close to the diffuse halo limiting estimate~\eqref{eq:cb_dh}, while at large $m_\chi$ (i.e., dark core regime) the baryonic compactness increases, and approaches the dark core limit result~\eqref{eq:cb_dc}. The same qualitative trend holds for both $M_\tl =1.4\,\msun$ and $M_\tl =2\,\msun$ sequence of solutions.

Now that we understand the behaviors of $R_\bb$ and $C_\bb$ across the \DANS{} solution space,
we now turn our attention to $\Delta R$.
Substituting Eqs.~\eqref{eq:cb_dc} and~\eqref{eq:cb_dh} into Eq.~\eqref{eq:ct_b6}, and then solving 
for $\Delta R$, reveals that diffuse halo configurations are expected to have a thicker crust 
(closer to the \PNS{} value) than dark core configurations. This behavior is confirmed in Fig.~\ref{fig:crust_ratio}.
In Appendix~\ref{app:ct_table}, we discuss the accuracy of the resulting expression for $\Delta R$;
see Fig.~\ref{fig:ct_relerr} therein. The main take away is the following: Eq.~\eqref{eq:ct_b6} reproduces our numerical results for the 
crust thickness at the percent level across our solution space and at sub-percent level in best-case scenarios.

\begin{figure}[t]
    \includegraphics[width=\columnwidth]{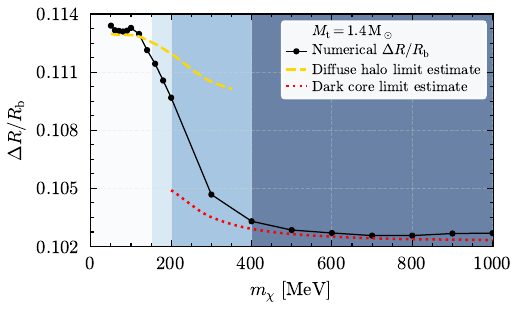}
    \caption{The crust thickness fraction as a function of $\mc$ for \DANS{} models with fixed $M_\tl=1.4\,\msun$, and dark matter mass fraction $f_\chi=0.05$.
    The black line denotes the numerical result obtained from the two-fluid TOV solutions. The dashed curve shows the diffuse-halo limit estimate using the baryonic compactness $C_\bb$ in Eq.~\eqref{eq:cb_dh}, while the dotted curve shows the dark-core limit estimate using Eq.~\eqref{eq:cb_dc}. Both estimates agree well with the numerical data in their regimes of validity. The shaded regions indicate different \DANS{} configurations, 
    classified by the halo-mass fraction $\delta_M$, as we did in Fig.~\ref{fig:crust_ratio}.}
    \label{fig:ct_frac_limits}
\end{figure}

It is interesting to study Eq.~\eqref{eq:ct_b6} in the two limiting baryonic compactness cases. We do so in Fig.~\ref{fig:ct_frac_limits}.
The dashed line denotes the diffuse halo limit, where the baryonic compactness is approximated by Eq.~\eqref{eq:cb_dh}, while the dotted line denotes the dark core limit, where the compactness is approximated by Eq.~\eqref{eq:cb_dc}. The black curve shows the numerical fractional crust thickness.
Reassuringly, for the $M_\tl=1.4\,\msun$ sequence shown here, the limiting case estimates reproduce well the numerical behavior within their corresponding regimes of validity. The diffuse halo estimate follows the numerical result closely at low $\mc$. Likewise, the dark core estimate agrees well at large $\mc$.
To illustrate the range of validity of these approximations, we extended the dashed and dotted lines slightly beyond their nominal regimes of validity. In these extended regions, the  two estimates
begin to deviate from the numerical results, showing that each expression is accurate mainly in the limit for which it was derived, as expected. These results are qualitatively the same for the $M_\tl=2\,\msun$ sequence of solutions.

\subsection{Why does the crust becomes thinner?}

\subsubsection{Pressure profile analysis}

\begin{figure}[h]
    \includegraphics[width=0.974\columnwidth]{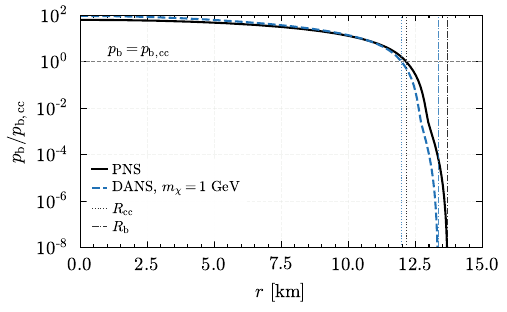}
    \caption{Baryonic pressure profiles normalized by the crust-core transition pressure, $p_\bb(r)/p_{\bb,\rm{cc}}$, for a \DANS{} solution with $M_\tl=1.4\,\msun$, $\mc=1\,\mathrm{GeV}$, and vanishing self-interaction, compared with a \PNS{} model with total mass $M_{\bb}=1.33\,\msun$. The horizontal dashed line marks $p_\bb=p_{\bb,\textrm{cc}}$, while the vertical lines indicate the crust-core radius $R_\textrm{cc}$ and baryonic surface radius $R_\bb$ for each model. The pressure gradient for the \DANS{} solution is larger near $R_\bb$ in comparison with the \PNS{} solution, thus resulting in a thinner crust.}
    \label{fig:pres_ratio}
\end{figure}

Regardless of how much the crust thickness of our \DANS{} solutions deviated
from its value for a \PNS{} reference solution, one pattern emerged from Figs.~\ref{fig:crust_ratio}
and~\ref{fig:crust_ratio_y}: the crust
was always thinner for the \DANS{}.
The reason why the crust becomes thinner can be understood by analyzing the baryonic pressure profile, $p_\bb(r)$.
In Fig.~\ref{fig:pres_ratio}, we compare the pressure profiles of an illustrative
\DANS{} solution with $M_\tl=1.4\,\msun$ and $m_\chi=1\,\mathrm{GeV}$ to its corresponding
reference \PNS{} solution of mass $1.33\,\msun$.
The \DANS{} pressure profile (dashed line) 
decreases more rapidly in the crust, so the interval between $R_\mathrm{cc}$ and $R_\bb$ is smaller than 
for
the \PNS{} pressure profile (solid line). As a consequence, $\Delta R$ is smaller for the \DANS{}.
This inward shift of the pressure profile appears for all \DANS{} solutions we considered,
and becomes more pronounced as $\mc$ increases. 
This behavior gives a direct illustration of the progressive crust-thickness reduction shown in Fig.~\ref{fig:crust_ratio}. 
We observed a similar behavior for other \DANS solutions.

\subsubsection{Approximate formula analysis}

We now study the behavior of $\Delta R$ as a function of $\mc$.
By taking the derivative of Eq.~\eqref{eq:ct_b6} with respect to $\mc$, we find
\begin{align}
    \frac{\dd}{\dd m_\chi}
    \left(
    \frac{\Delta R}{R_\bb}
    \right)
    &=
    \frac{\dd C_\bb}{\dd \mc}
    \frac{\dd}{\dd C_\bb}
    \left(
    \frac{\Delta R}{R_\bb}
    \right)
    \nonumber \\
    &=
    -\frac{\alpha}
    {[
    C_\bb+ \alpha \, (1-2C_\bb)
    ]^2}
    \frac{\dd C_\bb}{\dd m_\chi}.
\end{align}
Since $\alpha>0$, the fractional crust thickness decreases whenever $\dd C_\bb/ \dd m_\chi>0$, as shown in Fig.~\ref{fig:Cb}.
For the crust thickness, we write $\Delta R_\bb = R_\bb (\Delta R / R_\bb)$,
which yields,
\begin{align} \label{eq:dDr_dmchi}
\begin{aligned}
    \frac{\dd \Delta R}{\dd m_\chi}
    &=
    -
    R_\bb
    \, \frac{\alpha}
    {[
    C_\bb + \alpha\,(1-2C_\bb)
    ]^2}
    \frac{\dd C_\bb}{\dd m_\chi}
    \\
    &\quad +
    \frac{\Delta R}{R_\bb}
    \frac{\dd R_\bb}{\dd m_\chi}.
\end{aligned}
\end{align}
As shown in Figs.~\ref{fig:rb_pns_ratio} and \ref{fig:Cb}, our numerical
results satisfy $\dd R_\bb/ \dd m_\chi<0$ and $\dd C_\bb/ \dd m_\chi>0$, 
for fixed $M_\tl$ and $f_\chi$. Both terms in Eq.~\eqref{eq:dDr_dmchi} are therefore negative, 
implying,
\begin{equation}
    {\dd \Delta R} / {\dd m_\chi}<0 .
\end{equation}
This result is consistent with the crust thickness reduction we saw in Fig.~\ref{fig:crust_ratio}.

Having understood how the presence of dark matter changes $\Delta R$, we now proceed to study
how it affects the torsional crust oscillations.

\section{Torsional oscillations in the Cowling approximation} \label{sec:tors}

In this section we begin our study of how the presence of dark matter affects the
neutron-star torsional oscillations.
We start by deriving the master equation that describes these oscillations
in Sec.~\ref{sec:tors:master_eq}.
We then present how we compute the crust's shear modulus in
Sec.~\ref{sec:tors:shear}, and how we compute the oscillation
frequencies in Sec.~\ref{sec:tors:num}.
We present our numerical results in Sec.~\ref{sec:results}.

\subsection{Derivation of the master equation} \label{sec:tors:master_eq}

We consider torsional oscillations of the crust, adopting
the relativistic Cowling approximation, in which metric perturbations
are neglected~\cite{McDermott:1983ApJ,Finn:1988MNRAS}.
This approximation is well justified for torsional modes because they are
nonradial oscillations that do not significantly perturb
the spacetime geometry~\cite{Schumaker:1983MNRAS.203..457S};
they are odd-parity modes in the language of Ref.~\cite{thorne:1967ApJ}.

These modes involve no radial or polar matter displacement, and do not
perturb density and pressure to leading order.
The only nonzero component of the Lagrangian displacement vector
is the azimuthal component:
\begin{equation} \label{eq:xi_phi}
\xi^\phi(t,r,\theta)
= \frac{1}{\sin\theta}\, Y_\ell(t,r)\,
\partial_\theta P_\ell(\cos\theta),
\end{equation}
where $P_\ell$ is the Legendre polynomial of degree $\ell$ and $Y_\ell$ encodes the temporal and radial dependence of the mode;
because of spherical symmetry we consider $m=0$ modes only.
Since (i) we are interested in crustal oscillations (described by baryonic
matter) and (ii) baryonic and dark-matter fluids interact gravitationally only,
we have:
\begin{equation}
\delta u^\mu_\dd=0,
\qand
\delta T^{\mu\nu}_\dd=0,
\end{equation}
that is, only the baryonic fluid contributes to the dynamical perturbation
equations. The effect of dark matter is constrained
to be through the background (unperturbed) \DANS{} solution only.

The background baryonic four-velocity is
\begin{equation}
u^\mu_\bb = \ee^{-\Phi}(1,0,0,0).
\end{equation}
Using Eq.~\eqref{eq:xi_phi}, we find that the only nonzero component of the perturbed four-velocity,
\begin{equation}
\delta u^\mu = \delta u^\mu_\bb = u^{t} \, \partial_t \xi^{i},
\end{equation}
is:
\begin{equation} \label{eq:delta_u_phi}
\delta u^\phi
= \ee^{-\Phi} \, (\sin\theta)^{-1} \,
\partial_t Y_\ell\,
\partial_\theta P_\ell,
\end{equation}
while the dark matter four-velocity stays unperturbed.

The baryonic energy-momentum tensor in the crust is given by the sum of
a perfect-fluid and shear parts,
\begin{equation}
T^{\mu\nu}_\bb= T^{\mu\nu}_{\rm pf} + T^{\mu\nu}_{\rm shear},
\end{equation}
where
\begin{equation}
T^{\mu\nu}_{\rm pf}=(\varepsilon_\bb + p_\bb)u^\mu_\bb u^\nu_\bb + p_\bb g^{\mu\nu}
\end{equation}
and
\begin{equation}
T^{\mu\nu}_{\rm shear}=-2\mu S^{\mu\nu}.
\end{equation}
Here, $\mu$ is the shear modulus and $S^{\mu\nu}$ is the shear tensor
Since variations of the energy density and pressure vanish for axial perturbations,
\begin{equation}
\delta \varepsilon_\bb = 0,
\qand
\delta p_\bb = 0,
\end{equation}
and use the Cowling approximation, the perturbation to the perfect fluid energy-momentum tensor
arises only from the perturbed four-velocity,
\begin{equation}
\delta T^{\mu\nu}_{\rm pf}
=
(\varepsilon_\bb+p_\bb)\left(u_\bb^\mu\,\delta u_\bb^\nu+\delta u_\bb^\mu\,u_\bb^\nu\right),
\label{eq:deltaT_pf_axial}
\end{equation}
while the perturbed shear tensor is
\begin{equation}
\delta T^{\mu\nu}_{\rm shear}
=
-2\mu\,\delta S^{\mu\nu},
\label{eq:deltaT_shear_axial}
\end{equation}
and, therefore,
\begin{equation}
\delta T^{\mu\nu}_{\bb}
=
\delta T^{\mu\nu}_{\rm pf}
+
\delta T^{\mu\nu}_{\rm shear}.
\label{eq:deltaT_total_axial}
\end{equation}

From Eq.~\eqref{eq:delta_u_phi} we already know Eq.~\eqref{eq:deltaT_pf_axial}.
To proceed, we must calculate the components of perturbed shear tensor.
We first introduce the projection operator orthogonal to the baryonic fluid four-velocity,
\begin{equation}
P_{\mu\nu} = g_{\mu\nu}+u^\bb_{\mu}u^\bb_{\nu},
\quad
P^\mu{}_{\alpha} \, u_\bb^\alpha = 0,
\quad
P_{\mu\nu} = P_{\nu\mu},
\label{eq:proj_def}
\end{equation}
in terms of which we define the rate of shear as
\begin{equation}
\sigma_{\mu\nu}
=
\tfrac{1}{2} (
P_{\mu}{}^{\alpha} \, \nabla_{\alpha} u^\bb_{\nu}
+
P_{\nu}{}^{\alpha} \, \nabla_{\alpha}u^\bb_{\mu}
)
-\tfrac{1}{3}P_{\mu\nu} \, \nabla_{\alpha}u_\bb^{\alpha}.
\label{eq:sigma_def}
\end{equation}
Note that $\sigma_{\mu\nu} = \sigma_{\nu\mu}$~\cite{Schumaker:1983MNRAS.203..457S}.

Since the unperturbed star is shear-free, $\sigma_{\mu\nu}=0$, the shear contribution to the perturbed energy-momentum tensor depends only on $\delta\sigma_{\mu\nu}$. By linearizing Eq.~\eqref{eq:sigma_def} and working in the Cowling approximation, $\delta g_{\mu\nu}=0$, we find
\begin{align}
\begin{aligned}
\delta\sigma_{\mu\nu}
&=
\tfrac{1}{2} \, (
\delta P_{\mu}{}^{\alpha} \, \nabla_{\alpha}u^\bb_{\nu}
+
\delta P_{\nu}{}^{\alpha} \, \nabla_{\alpha}u^\bb_{\mu}
)
\\
&\quad
+
\tfrac{1}{2} \, (
P_{\mu}{}^{\alpha} \, \nabla_{\alpha}\delta u^\bb_{\nu}
+
P_{\nu}{}^{\alpha} \, \nabla_{\alpha}\delta u^\bb_{\mu}
)
\\
&\quad
-\tfrac{1}{3} \, ( \delta P_{\mu\nu} \, \nabla_{\alpha}u_\bb^{\alpha} + P_{\mu\nu} \, \nabla_{\alpha}\delta u_\bb^{\alpha} ),
\end{aligned}\label{eq:delta_sigma}
\end{align}
where the perturbed projection tensor is:
\begin{equation}
\delta P_{\mu\nu}
=
u^\bb_\mu\,\delta u^\bb_\nu
+
u^\bb_\nu\,\delta u^\bb_\mu .
\label{eq:deltaP_def}
\end{equation}
Because the only nonzero background four-velocity component is $u^\bb_t = -\exp\Phi$
and the only nonzero perturbed four-velocity component is
$\delta u^\bb_\phi$,
the nonzero components of $\delta P_{\mu\nu}$ are:
\begin{equation}
\delta P_{t\phi}=\delta P_{\phi t}
=
-\,r^2\sin\theta\,
\partial_t Y\,
\partial_\theta P_\ell.
\label{eq:deltaP_0phi}
\end{equation}
Substituting this result in Eq.~\eqref{eq:delta_sigma},
we obtain the nonzero components of the perturbed rate of shear tensor:
\begin{subequations} \label{eq:delta_sigma_components}
\begin{align}
&\delta\sigma_{r\phi}
=
\delta\sigma_{\phi r}
=
-\tfrac{1}{2} \, \ee^{-\Phi} \, r^2\sin^2\theta\,
\partial_\theta P_\ell \,
\partial_{tr} Y,
\label{eq:delta_sigma_rphi_explicit}
\\
&\delta\sigma_{\theta\phi}
= \delta\sigma_{\phi\theta}
=
\tfrac{1}{2} \, \ee^{-\Phi} \, r^2\sin^3\theta\,
\partial^2_\theta P_\ell \,
\partial_t Y.
\label{eq:delta_sigma_comps}
\end{align}
\end{subequations}

Having found $\delta P_{\mu\nu}$ and $\delta\sigma_{\mu\nu}$,
we can obtain the perturbed shear tensor, $\delta S_{\mu\nu}$, as follows.
The rate of shear tensor is defined as the Lie derivative of the
shear tensor along the baryonic fluid wordline~\cite{Carter:1977qf}. %
To leading order in perturbation we have
\begin{equation}
\delta\sigma_{\mu\nu}
=
\mathcal{L}_{u_\bb}[\delta S_{\mu\nu}].
\label{eq:delta_sigma_Lie_deltaS}
\end{equation}
Linearizing the Lie derivative, we obtain
\begin{align}
\mathcal{L}_{u_\bb}[\delta S_{\mu\nu}]
&=
u^\alpha_\bb \nabla_\alpha \delta S_{\mu\nu,\bb}
+
\delta S_{\alpha\nu,\bb}\nabla_\mu u^\alpha_\bb
\nonumber \\
&\quad + \delta S_{\mu\alpha,\bb}\nabla_\nu u^\alpha_\bb,
\end{align}
where the last two terms vanish because the background flow is static and $\delta S_{\mu\nu,\bb}$ has nonzero $\phi$-components only.
We then obtain:
\begin{equation}
\delta\sigma_{\mu\nu}
=
\ee^{-\Phi}\,\partial_t \delta S_{\mu\nu}.
\label{eq:sigma_S}
\end{equation}

We can now substitute Eq.~\eqref{eq:delta_sigma_components} in Eq.~\eqref{eq:sigma_S},
do one integrating in time, using as initial conditions that the unperturbed star is shear-free,
to obtain the nonzero components of the perturbed shear tensor:
\begin{subequations} \label{eq:delta_S_components}
\begin{align}
\delta S_{r\phi}
&=
\delta S_{\phi r}
=
-\tfrac{1}{2}\,r^2\sin^2\theta\,
\partial_r Y \,
\partial_\theta P_{\ell},
\\
\delta S_{\theta\phi}
&=
\delta S_{\phi\theta}
=
\tfrac{1}{2}\,r^2\sin^3\theta\,
Y \,
\partial_{\theta\theta} P_{\ell}.
\label{eq:deltaS_thphi}
\end{align}
\end{subequations}
Together, Eqs.~\eqref{eq:deltaT_pf_axial} and~\eqref{eq:delta_S_components}
completely determine the perturbed energy-momentum tensor~\eqref{eq:deltaT_total_axial}.

The master equation for the displacement function $Y_\ell$ follows
from the energy-momentum conservation of the baryonic fluid component.
In the Cowling approximation,
\begin{equation}
\nabla_\alpha \, \delta T^{\alpha}{}_{\phi,\bb}=0.
\label{eq:consv_phi}
\end{equation}
Carrying out the calculation, we obtain the wave equation
\begin{align}
- \frac{\ee^{2\Lambda-2\Phi}}{v^2_\bb} \,\partial_t^2 Y_{\ell}
&+ \partial_r^2 Y_{\ell}
+\left[\frac{4}{r} +\Phi' - \Lambda'+\frac{\mu'}{\mu}\right]\partial_r Y_{\ell}
\nonumber \\
&- \ee^{2\Lambda} \, \frac{(\ell+2)(\ell-1)}{r^2} \,Y_{\ell} = 0,
\label{eq:pde_master}
\end{align}
where a prime indicates a derivative with respect to $r$, and
$v_\bb$ is the wave propagation speed~\cite{Schumaker:1983MNRAS.203..457S},
\begin{equation} \label{eq:def_shear_speed}
    v_{\bb} = \left( \frac{\mu}{\varepsilon_\bb + p_\bb} \right)^{1/2}.
\end{equation}

It is convenient to work in the frequency domain. We assume that the fluid displacement has a harmonic time dependence,
\begin{equation}
    Y_\ell(t,r) = Y_\ell(r) \, \ee^{\ii \omega t},
\end{equation}
thereby reducing  Eq.~\eqref{eq:pde_master} to the form
\begin{align}
Y_\ell''
&+
\left[
\frac{4}{r}+\Phi'-\Lambda'+\frac{\mu'}{\mu}
\right] Y_\ell'
\nonumber \\
&+ \, \ee^{2\Lambda}\left[
\frac{\omega^2}{v_\bb^2} \, \ee^{-2\Phi}-
\frac{(\ell+2)(\ell-1)}{r^2}
\right] Y_\ell =0.
\label{eq:master}
\end{align}

For the purpose of numerical integration, we introduce $Z_{\ell} = {\dd Y_\ell}/{\dd r}$,
and write Eq.~\eqref{eq:master} as a pair of first-order coupled differential equations:
\begingroup
\allowdisplaybreaks
\begin{subequations} \label{eq:odesystem}
    \begin{align}
        \frac{\dd Y_\ell}{\dd r} &= Z_{\ell}, \\
        \frac{\dd Z_\ell}{\dd r} &=
        -\left[
        \frac{4}{r}+\Phi'-\Lambda'+\frac{\mu'}{\mu}
        \right] Z_\ell
        \nonumber\\
        &\quad \, -\ee^{2\Lambda}\left[
        \frac{\omega^2}{v_\bb^2}\,\ee^{-2\Phi}
        -\frac{(\ell+2)(\ell-1)}{r^2}
        \right] Y_\ell.
    \end{align}
\end{subequations}
\endgroup
We reiterate that while not explicit, dark matter enters Eq.~\eqref{eq:odesystem}
through the metric functions $\Phi$ and $\Lambda$, and their derivatives.
Equation~\eqref{eq:pde_master} [or its equivalent, Eq.~\eqref{eq:odesystem}] is the master
equation that describes the crustal torsional oscillations of a \DANS{} in the Cowling approximation.

To solve Eq.~\eqref{eq:odesystem} we need to impose boundary
conditions on the fluid displacement at the crust-core interface $R_{\cc\cc}$
and baryonic surface $R_\bb$. At these locations, we demand the
radial component of the stress tensor to vanish; this is the ``zero-torque'' condition~\cite{Schumaker:1983MNRAS.203..457S}:
\begin{equation} \label{eq:bcs}
    Z_{\ell}(r) =0
    \qquad \text{as} \qquad
    r \to R_{\cc\cc}^{+},\, R_\bb^{-}.
\end{equation}

With Eq.~\eqref{eq:bcs}, computing the oscillations frequencies
becomes an eigenvalue problem, which we solve using a shooting method.
The outcome are discrete angular frequency values $\omega = \omega_{\ell n}$, characterized
by the multipole number $\ell$ and the number $n$ of (radial) nodes of $Y_{\ell}$.
We will quote our results in terms of the frequency:
\begin{equation}
    {}_{\ell} t_{n} = \omega_{\ell n} / (2 \pi),
\end{equation}
following a notation common in the literature.

Before we can numerically integrate Eq.~\eqref{eq:odesystem}, we need a model to describe
the elastic properties of the solid crust, in particular its shear modulus $\mu$.
We discuss this next.

\subsection{The shear modulus} \label{sec:tors:shear}

Assuming the neutron-star crust to be a body-centered cubic
lattice, the shear modulus $\mu$ in the zero-temperature limit
can be approximated as:
\begin{equation} \label{eq:shear_no_es}
    \mu = 0.1194 \, n \, (Ze)^2 / a,
\end{equation}
where $n$ is the ion number density, $Z$ is the nuclei atomic number,
$e$ is the electron charge, and
$a^3 = 3 / (4 \pi  n)$
is the radius of the Wigner--Seitz cell containing one
nucleus~\cite{Ogata:1990PhRvA,Strohmayer:1991ApJ}.

It is often assumed that electrons are uniformly distributed
in the crust. A nonuniform distribution results in
electron screening which reduces the shear modulus.
Kobyakov and Pethick~\cite{Kobyakov:2013eta} showed that Eq.~\eqref{eq:shear_no_es} can be replaced
with:
\begin{equation} \label{eq:shear_es}
    \mu = 0.1194 \, n \, (1 - 0.010 \, Z^{2/3})\, (Ze)^2 / a,
\end{equation}
to account for electron screening.
The reduced shear modulus decreases the torsional oscillation frequencies.
For example, Sotani~\cite{Sotani:2014dua} found, using the
Kobyakov--Pethick~\cite{Kobyakov:2013eta} and Douchin--Hansel~\cite{Douchin:2001sv} crust
equations of state, that Eq.~\eqref{eq:shear_es} reduces the frequency of the fundamental
oscillation mode ${}_{2}t_{0}$ by 6\% in comparison to using Eq.~\eqref{eq:shear_no_es}.

In practice, we calculate $\mu$ using the composition information provided by
the \eos equation of state.
At each point in the crust, the equation of state gives the baryon number
density $n_\bb$, average mass number $A$, and average proton number ${Z_{\rm av}}$.
We compute the ion number density as $n = {n_\bb} / {A}$, from which we determine the Wigner--Seitz cell radius $a$.
We then substitute these quantities into
Eq.~\eqref{eq:shear_no_es} or Eq.~\eqref{eq:shear_es}, depending on whether
electron screening is included or not.

\subsection{Numerical integration} \label{sec:tors:num}

We now have all the necessary ingredients to calculate the torsional oscillation frequencies.
For a given equilibrium \DANS{} solution,
we know the background quantities $m_\bb$, $m_\dd$, $p_\bb$, $p_\dd$, $\Phi$, and shear modulus $\mu$.
We can then integrate Eq.~\eqref{eq:odesystem} in the crust, starting from $r = R_{\rm cc}$ and out to $R_\bb$,
for a chosen multipole number $\ell$ and frequency $\omega$.
Since Eq.~\eqref{eq:master} is linear and homogeneous, the
normalization of the eigenfunction is arbitrary. We choose
\begin{equation}
Y_\ell (R_{\rm cc})=1,
\end{equation}
while the inner zero-torque boundary condition gives
\begin{equation}
Z_\ell (R_{\rm cc})=0;
\end{equation}
these are our initial conditions.
The eigenfrequencies are the values of $\omega$ for which the outer zero-torque condition is satisfied, i.e., the frequencies for which
the function
\begin{equation} \label{eq:shooting_function}
S_{\ell}(\omega) = Z_\ell (R_\bb;\omega),
\end{equation}
vanishes.
At fixed $\ell$, Eq.~\eqref{eq:shooting_function} admits various roots corresponding to eigenfunctions
with a number $n$ of nodes; we call $n$ the overtone number.

\begin{figure*}[ht]
    \includegraphics[width=\columnwidth]{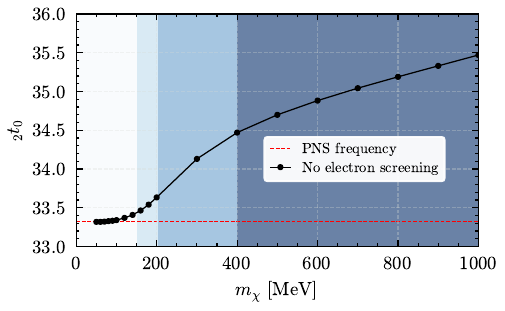}
    \includegraphics[width=\columnwidth]{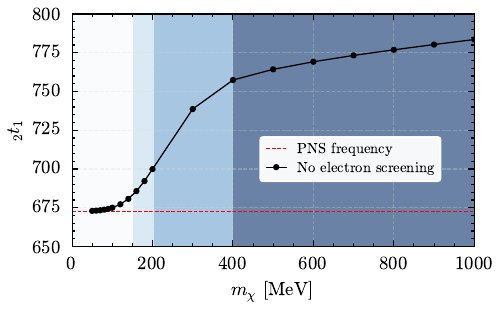}
    \caption{The quadrupolar torsional oscillations frequencies for \DANS{} solutions with fixed total mass $M_\tl=1.4\,\msun$ and dark-matter mass fraction $f_\chi = 0.05$.
    All calculations 
    were done without the inclusion of electron screening effects in the shear modulus. We show the fundamental mode, ${}_{2}t_{0}$, in the left panel and the first overtone, ${}_{2}t_{1}$, in the right panel. In both panels, we indicate with the dashed horizontal line the corresponding values of these frequencies for a comparable \PNS{} solution; see Table~\ref{tab:pns_freqs}.
    We see that the frequencies increase as a function $\mc$ in both panels. This behavior is consistent with the decrease in the crust thickness shown in Fig.~\ref{fig:crust_ratio}.
    As in previous figures, the shaded regions denote different \DANS{} configurations; see Fig.~\ref{fig:crust_ratio}.
    }
    \label{fig:mt14_nes}
\end{figure*}

\section{Torsional oscillation frequencies of dark-matter admixed neutron stars} \label{sec:results}

We now present our results for the torsional oscillation frequencies of \DANS{}.
As in Sec.~\ref{sec:dans}, we study two sequences of \DANS solutions parametrized by the
dark-matter particle mass $\mc$, dark-matter fraction,
$f_\chi = 0.05$, and total mass $M_\tl = 1.4\,\msun$ or $2\,\msun$; recall that we 
are neglecting the self-interaction.
For each solution along these sequences, we calculated the fundamental, ${}_{2}t_{0}$, and
first overtone, ${}_{2}t_{1}$, frequencies of the torsional modes following the
recipe outlined in Sec.~\ref{sec:tors:num}.
In this section, we neglect electron screening in the shear modulus.
For the same reason 
given
in Sec.~\ref{sec:dans:crust}, we first did this same
calculation for two \PNS{}s of total masses $1.33\,\msun$ and $1.9\,\msun$.
We summarize 
our
results in Table~\ref{tab:pns_freqs}.
These results will serve as benchmarks against which we will compare our results
for \DANS{} solutions with $M_\tl=1.4\,\msun$ and $M_\tl=2\,\msun$.

\begin{table}[t]
\caption{Properties of the two reference \PNS solutions. We show their
total mass $M$, radius $R$, crust thickness $\Delta R$, and fundamental and first overtone
oscillation frequencies of the quadrupole mode, ${}_{2} t_{n}$. In the last two columns,
the first (second) entry corresponds to the frequency without (with) electron screening effect added
to the shear modulus.}
\begin{tabular}{ c c c c c}
\arrayrulecolor{Black}
\hline \hline
$M$\,[$\msun$] & $R$\,[km] & $\Delta R$\,[km] & ${}_{2}t_{0}$\,[Hz] & ${}_{2}t_{1}$\,[Hz]  \\
\hline
1.33 & 13.702 & 1.556 & 33.317 (31.075) & 672.681 (631.930) \\
1.90 & 14.033 & 1.004 & 34.359 (32.042) & 1033.453 (970.619) \\
\hline \hline
\end{tabular}
\label{tab:pns_freqs}
\end{table}

Figure~\ref{fig:mt14_nes} shows the torsional oscillation frequencies for the fundamental mode (left panel)
and first overtone (right panel) for \DANS{} solutions with fixed total mass $M_\tl = 1.4\,\msun$.
In both panels, the dashed line represents the frequency of the corresponding \PNS{} model, across the full range of $\mc$
--- it provides a reference against which the \DANS{} results can be compared to.
We see that the frequency of the $\tfdm$ mode is generally higher for a \DANS{} in comparison to the reference \PNS{} case.
For small values of $\mc$, the deviation is small and the frequencies are close to the \PNS{} value.
For example, for $\mc = 50$~MeV, the difference between the two frequencies is $10^{-3}$~Hz.
As we increase $\mc$, the higher the frequencies become.
This trend is especially evident for compact-halo and dark-core configurations.
In the right panel of Fig.~\ref{fig:mt14_nes} we show the first overtone frequency ${}_{2}t_{1}$ for the same sequence of \DANS{} solutions.
The qualitative trend remains the same: the frequency increases with $\mc$ and departs more significantly
from the \PNS{} value in the compact halo and dark core regimes. Quantitatively, the effect is stronger for the first overtone than for the fundamental mode.

\begin{figure}[b]
    \includegraphics[width=\columnwidth]{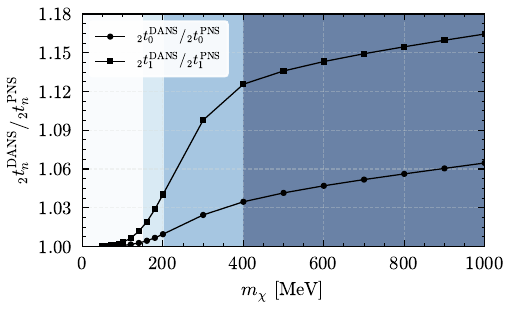}
    \caption{
    The ratio between fundamental mode (circles) and first overtone (squares) 
    torsional oscillation frequency ratios for 
    \DANS{} models with $M_\tl=1.4\,\msun$ relative to 
    a comparable \PNS{} model
    as functions of $\mc$. 
    The ratio between the frequencies increases monotonically with $\mc$, reaching maximum values 
    of approximately 6.5\% for fundamental mode and 16.4\% for the first overtone as $\mc$ 
    approaches 1~GeV.}
    \label{fig:mt14_nes_ratio}
\end{figure}

\begin{figure*}[t]
    \includegraphics[width=\columnwidth]{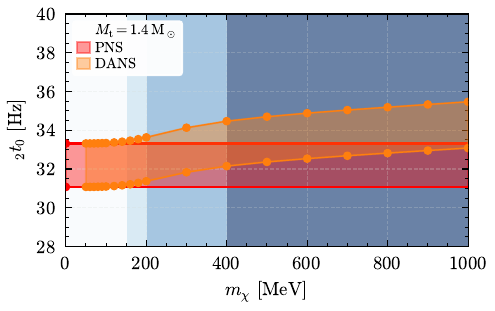}
    \includegraphics[width=\columnwidth]{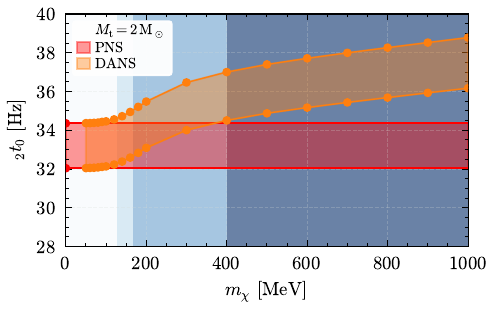}
    \caption{
    The fundamental torsional oscillation frequencies ${}_{2}t_{0}$ for \DANS{} models of fixed total mass $M_\tl = 1.4\,\msun$ (left panel) and $M_\tl = 2\,\msun$ (right panel), and dark-matter mass fraction $f_\chi = 0.05$. The curved band corresponds to frequencies for \DANS{} including (lower end) or not (upper end) electron screening effects on the shear modulus of the crust. The horizontal band is the equivalent case for a comparable \PNS{}. The degeneracy between electron-screening and dark-matter effects on ${}_{2}t_{0}$ is broken when the two bands do not overlap. For $M_\tl = 1.4\,\msun$ stars the two effects are always degenerate, but for $M_\tl = 2\,\msun$ stars the degeneracy is broken when the \DANS{s} are in the dark-core regime. 
    The vertical shaded regions denote different \DANS{} configurations; see Fig.~\ref{fig:crust_ratio}.}
    \label{fig:0t2_band}
\end{figure*}

Our findings in Fig.~\ref{fig:mt14_nes} are consistent with our study of the crust thickness in Sec.~\ref{sec:dans:crust}; cf. Fig.~\ref{fig:crust_ratio}, in particular.
Therein, we showed that the \DANS{} solutions have a thinner crust relative to their reference \PNS{}.
By making the crust thinner, the presence of dark matter increases the mode frequencies,
since they scale inversely with the crust thickness~\cite{Samuelsson:2006tt}.
The largest gradient in $\Delta R$ ocurred for compact-halo configurations;
this is mirrored in where ${}_{2}t_{n}$ changes most rapidly as a function of
$\mc$ in Fig.~\ref{fig:mt14_nes}.

To quantify the magnitude of the shifts, in Fig.~\ref{fig:mt14_nes_ratio} we plot
the ratio between our \DANS{} and \PNS{} results for both modes.
For the fundamental mode, we see an increase of about $6.4\%$ at $\mc = 1$~GeV.
For the first overtone, the increase reaches about $16.4\%$,
hence being more sensitive to the presence of dark matter.

Our study so far has isolated the effects of dark matter on the torsional frequencies from the uncertainties in modeling the baryonic physics of the crust.
In particular, the presence of dark matter \emph{increases} the frequencies.
How is this outcome impacted by the inclusion of electron screening that softens the shear modulus, thus \emph{decreasing} the frequencies? We investigate the competition between these two effects next.

\section{Degeneracy between dark-matter and electron screening effects} \label{sec:results_es}

\begin{figure*}[t]
    \includegraphics[width=\columnwidth]{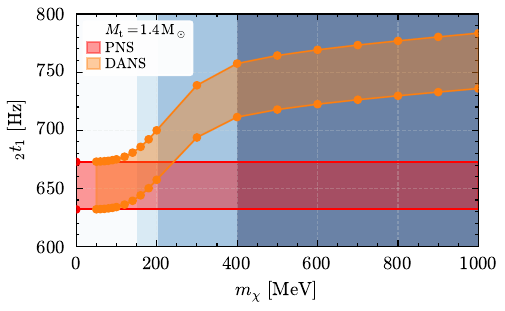}
    \includegraphics[width=\columnwidth]{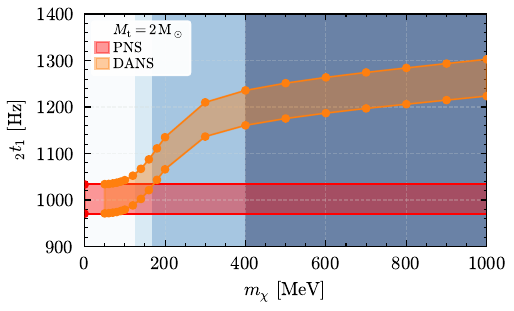}
    \caption{Similar to Fig.~\ref{fig:1t2_band}, but for the first overtone torsional oscillation frequencies, ${}_{2}t_{1}$. Contrary to the fundamental frequency, we see that the degeneracy between electron-screening and dark-matter effects is broken for both $M_\tl = 1.4\,\msun$ and $M_\tl = 2\,\msun$ stars, as long as $\mc$ is large enough for the \DANS{} solutions to be of the compact-halo or dark-core types. Moreover, the deviations from the \PNS{} frequencies are larger for the first overtone than for the fundamental frequency.}
    \label{fig:1t2_band}
\end{figure*}

To study the combined effects of electron screening and presence of dark matter on the torsional oscillation spectrum, we repeated our calculations of Sec.~\ref{sec:results},
but now using Eq.~\eqref{eq:shear_es} for the shear modulus.

In Fig.~\ref{fig:0t2_band}, we compare the fundamental mode frequencies, ${}_{2}t_{0}$, computed with and without electron screening.
The left (right) panel corresponds to \DANS{} solutions with fixed total mass $M_\tl = 1.4\,\msun$ ($M_\tl = 2\,\msun$).
The upper orange circles denote the results without electron screening, while the lower orange circles denote those with electron screening. The region between these two curves is labeled ``\DANS{}'' in the figure.
Since a \PNS{} contains no dark matter, its frequencies do not depend on $\mc$. As we did in Fig.~\ref{fig:mt14_nes}, we plot the corresponding \PNS{} frequencies with and without electron screening as horizontal lines and shade the region between them. This gives a reference band, labeled ``\PNS{},'' against which we compare our \DANS{} band to. 
These two bands can be interpreted as the frequency range bracketed by the two shear modulus prescriptions we consider.
In both panels, we see that the two \DANS{} curves are nearly parallel to each other.
Therefore, electron screening mainly decreases the fundamental mode frequency, while leaving the dependence on $\mc$ unchanged.
For the lower mass case, $M_\tl = 1.4\,\msun$, the two bands overlap in the entire range of $\mc$. This means
that the two effects (presence of dark matter and electron screening) are degenerate. This degeneracy is weaker when $\mc$ approached 1~GeV, but is always present. For the higher mass case, $M_\tl = 2\,\msun$, the situation is different. We see that the two bands no longer overlap when $\mc \gtrsim 400$~MeV, which corresponds to the dark core regime: the degeneracy is broken therein.

In Fig.~\ref{fig:1t2_band}, we repeat the same analysis but now for the first overtone, ${}_{2}t_{1}$.
Compared with the fundamental mode, the degeneracy between \DANS{} and \PNS{} bands is now weaker.
For the lower-mass case, $M_\tl = 1.4\,\msun$, the \DANS{} band branches off
the \PNS{} band in the compact-halo and dark-core regimes. For the higher-mass case, $M_\tl = 2\,\msun$, this separation is even more pronounced.

In summary, to tell apart the \DANS{} and \PNS{} models using the fundamental frequency ${}_{2}t_{0}$, 
we need a relatively high stellar mass and large dark matter particle mass, e.g., $\mc \gtrsim 400$~MeV for $M_\tl = 2\,\msun$.
The first overtone, ${}_{2}t_{1}$, is more sensitive to the presence of dark matter: the \DANS{} and \PNS{} bands become distinguishable for $\mc \gtrsim 250$~MeV at $M_\tl = 1.4\,\msun$, and for $\mc \gtrsim 200$~MeV at $M_\tl = 2\,\msun$.

We tabulated all torsional oscillation frequencies we computed 
in Secs.~\ref{sec:results} and~\ref{sec:results_es} in Appendix~\ref{app:tor_num_results}.

\section{Conclusions} \label{sec:discussions}

We investigated how a gravitationally coupled dark-matter fluid modifies the structure of the crust and the quadrupole torsional oscillation spectrum of \DANS{s}. 

In Sec.~\ref{sec:dans}, we constructed equilibrium \DANS{} solutions in the two-fluid formalism, using the \eos equation of state for baryonic matter and a repulsive self-interacting fermionic equation of state for dark matter. By solving the two-fluid TOV equations, we obtained the background stellar quantities required for the rest of this work, including the baryonic radius, dark matter radius, crust-core transition radius, and masses.
Through this work, we studied two sequences of \DANS{} solutions with fixed dark-matter 
mass fraction $f_\chi = 0.05$ and total gravitational masses of either $M_\tl = 1.4\,\msun$ or $2\,\msun$.

With these equilibrium solutions in hand, in Sec.~\ref{sec:crust} we studied how dark matter affects the baryonic crust thickness $\Delta R$, both numerically and analytically. 
We showed that heavier dark matter particles shift the \DANS{} solutions from halo to core-concentrated configurations. As a result, a larger fraction of the dark matter fluid is enclosed within the baryonic radius. This reduces the baryonic radius and increases the baryonic compactness; see Figs.~\ref{fig:rb_pns_ratio} and~\ref{fig:Cb}, respectively.
These two effects together, decrease the crust thickness relative to a comparable \PNS{} solution, as shown in Fig.~\ref{fig:crust_ratio}. We also found that at 
fixed dark-matter particle mass $\mc$, stronger repulsive self-interaction $y$ drives the 
solutions
away from dark-core toward halo configurations; see Fig.~\ref{fig:crust_ratio_y}. 
This reduces the deviations from the crust thickness compared to \PNS{s}.
Qualitatively, we explain these results as follow.
In the diffuse halo regime (small $\mc$), only a small fraction of the dark matter fluid contributes to the gravitational field inside the baryonic surface $R_\bb$. Hence, the baryonic compactness remains close to the \PNS{} value and the crust thickness barely changes.
In the dark core limit (large $\mc$), the baryonic crust responds to the full dark-matter fluid mass enclosed in $R_\bb$. This leads to a larger baryonic compactness and a thinner crust.

In Sec.~\ref{sec:tors}, we derived the perturbation equation governing the torsional oscillations in the relativistic Cowling approximation; see Eq.~\eqref{eq:master}. 
We also reviewed two ways to model the shear modulus of the neutron star crust, which includes or not electron screening effects.
We solved this equation numerically in Secs.~\ref{sec:results} and~\ref{sec:results_es}.
First, to isolate the effects of dark matter, we studied the torsional oscillation spectrum 
neglecting electron-screening effects in the shear modulus.
We found that the presence of dark matter increases the oscillation frequency for both the fundamental and first overtone modes. This increase is largest for heavier dark matter particles, consistent with the crust thickness becoming thinner for more core-concentrated dark matter distributions.
Next, 
we 
studied
the combined effects of electron screening and dark matter on the frequencies. This allowed us to examine the potential degeneracy in the oscillation spectrum caused by 
crust
microphysics and dark matter effects.
Our main results were summarized in Figs.~\ref{fig:0t2_band} and~\ref{fig:1t2_band}.
We found that electron screening and dark matter effects on the fundamental mode are 
degenerate unless we have a high-mass \DANS{} and heavy dark matter particle mass, approximately above $400$~MeV. 
However, for the first overtone, we found that this degeneracy is broken 
for both low- and high-mass \DANS{s}
as long as 
$\mc \gtrsim 200$~MeV.

Our results suggest that should torsional oscillations be identified, e.g., from the QPOs in the decaying tails of giant magnetar flares, they could provide a seismic probe of a dark matter core within neutron stars.
Our results also suggest that the crust can carry information about the spatial distribution of dark matter in \DANS{s}~\cite{Miao:2022rqj,Shawqi:2024jmk,Karkevandi:2024vov,Liu:2024swd}, complementing constraints from mass-radius measurements~\cite{Leung:2011zz,Das:2020ecp,Leung:2022wcf,Kumar:2025ytm}, tidal deformabilities~\cite{Karkevandi:2021ygv,Sagun:2022ezx,Kumar:2025ytm}, and fluid oscillation modes~\cite{Leung:2012vea, Routaray:2022utr,Sotani:2025hzb}.

Our work can be extended in a number of directions. 
For example, we could broaden the set of equations of state and stellar models we studied here. In particular, we used a single baryonic equation of state and 
dark matter model. Carrying out a systematic survey over 
a larger catalog of baryonic equations of state, dark matter models, stellar masses, and dark matter fractions would clarify which features of the torsional oscillation spectrum found here are generic. {We could also study higher multipole $\ell$ and overtone numbers $n$ modes.

It would also be important to incorporate other crust microphysics effects beyond electron screening. For example, nuclear pasta phases near the crust-core interface can modify the effective elastic region and the shear modulus profile~\cite{Sotani:2012qc,Sotani:2018tdr}. These works showed that ignoring pasta elasticity tends to underestimate torsional frequencies, and that including pasta elasticity can increase the fundamental mode when a finite pasta region is present.
They also found that the first overtone is especially sensitive to the crust thickness and to the crust microphysics near the inner boundary of the elastic region.
Therefore, including pasta elasticity in \DANS{} models is a natural next step for determining whether the dark matter-induced frequency shifts found here remain distinguishable from uncertainties in the crust microphysics.

It would also be important to go beyond the relativistic Cowling approximation. Although this is a good approximation for describing the torsional oscillations modes, we anticipate that a full calculation including metric perturbation as done by Schumaker and Thorne~\cite{Schumaker:1983MNRAS.203..457S}, would couple both baryonic and dark-matter fluid perturbations. As a consequence, we expect the appearance of a additional dark-matter oscillation mode in the crust region. It would be interesting to study if 
this new mode leaves
any observable imprints on the torsional oscillation spectrum.

Finally, of course, including magnetic fields~\cite{Messios:2001br,Glampedakis:2006apa,Sotani:2006at,deSouza:2018ope,Burnaz:2025tza}, rotation~\cite{Vavoulidis:2007ui,Vavoulidis:2007cs}, and extending the analysis to polar perturbations~\cite{1990MNRAS.245...82F,Yoshida:2002vd, Kruger:2024fxn,Sotani:2024hsm} are important steps toward a more complete asteroseismology picture of \DANS{s}. Evidently there remains much to be done.

\section*{Acknowledgments}
We thank Patrizio Di Lorenzo, Jay Kalinani, Sagnik Saha, Spyros Thomopoulos, and Helvi Witek for discussions. 
J.Z. acknowledges the American Physical Society (APS) Division of Astrophysics and the Department of Physics of the University of Illinois Urbana-Champaign for financial support.
H.O.S thanks Jonas P. Pereira and the Universidade de Bras\'ilia for the kind hospitality during the final writing stages of this work.

\appendix

\section{Approximate analytical formulas for the crust thickness} \label{app:crust_thickness}

Here, we derive the approximate analytical formula~\eqref{eq:ct_b6} for the crust thickness, following Refs.~\cite{Samuelsson:2006tt,Baym:2017whm}.
Their derivations apply to a single-fluid neutron star and provides a simple analytic relation between the fractional crust thickness and the stellar compactness.

The derivation relies on three simplifying assumptions. First, the crust is assumed to be thin compared with the stellar radius, so that metric and thermodynamic quantities vary only slightly across the crust. 
Second, the crust is assumed to be light so that the mass enclosed in the crust is neglected, and the total gravitational mass is approximately constant throughout this region and can be replaced by its value at the stellar surface.\footnote{These two assumptions are well motivated for astrophysically relevant neutron stars. For realistic equations of state and masses $M_\tl > 1\,\msun$ the crust mass $M_\tl-m(R_\mathrm{cc})$ is less than $3\%$ of the total gravitational mass $M_\tl$, while the crust thickness $\Delta R$ does not exceed $15\%$ of the stellar radius~\cite{Chamel:2008ca}.}
Third, the fluid in the crust is treated as barotropic, which allows the pressure equation to be rewritten in terms of the baryonic chemical potential, $\mu_\bb$.
Under these three assumptions, we can obtain an analytic estimate for the crust thickness 
that depends mainly on the compactness and on the ratio of chemical potentials at the 
crust-core interface, $\mu_{\rm cc} = \mu_\bb(R_{\rm cc})$, and at the surface, 
$\mu_{\rm s}  = \mu_\bb(R_\bb)$.

Using Eq.~\eqref{eq:tov_pb} and the relation for matter at zero temperature in full chemical equilibrium
\begin{equation} \label{eq:zero_tem}
    \dd p_\bb = n_\bb\, \dd \mu_\bb,
\end{equation}
together with
\begin{equation}
    \varepsilon_\bb + p_\bb = \mu_\bb n_\bb,
\end{equation}
we obtain
\begin{equation}
    {\dd \mu_\bb} / {\dd r} = -\mu_\bb \, ({\dd\Phi} / {\dd r}),
\end{equation}
or equivalently
\begin{equation}
    {\dd \ln \mu_\bb} / {\dd r} = - {\dd \Phi} / {\dd r}.
\end{equation}
Integrating this relation across the crust , from the crust-core interface radius $R_\cc$ to the baryonic surface radius $R_\bb$, and defining $\alpha=\ln(\mu_{\rm cc}/\mu_\su)$ gives
\begin{equation}
    \alpha =
    \int_{R_{\textrm{cc}}}^{R_\bb} \Phi'(r) \, \dd r.
\end{equation}
Assuming that the crust is thin and light, and that the baryonic pressure in Eq.~\eqref{eq:tov_pb} is small, $4\pi r^3 p_\tl \ll m_\tl(r)$, we can neglect the variation of the enclosed mass across the crust and set $m_\tl(r)\approx M_\tl(R_\bb)$. Since $r\approx R_\bb$ throughout the thin crust, we further approximate $M_\tl(R_\bb)/r\approx C_\bb$, where $C_\bb$ is the baryonic compactness defined in Eq.~\eqref{eq:def_compactness}.
Then,
\begin{equation}
    \Phi'(r)
    \approx
    \frac{M_\tl(R_\bb)}{r\left[r-2M_\tl(R_\bb)\right]}
    \approx
    \frac{C_\bb}{1-2C_\bb}\frac{1}{r}.
\end{equation}
Integration gives 
\begin{equation}
    \alpha
    \approx
    \frac{C_\bb}{1-2C_\bb}
    \int_{R_{\textrm{cc}}}^{R_\bb}
    \frac{\dd r}{r}
    =
    \frac{C_\bb}{1-2C_\bb}
    \ln\left(\frac{R_\bb}{R_{\textrm{cc}}}\right).
\end{equation}
Defining
\begin{equation}
    x = 
    {\Delta R} / {R_\bb}
    = 
    (R_\bb-R_{\textrm{cc}}) / R_\bb,
\end{equation}
so that $R_{\textrm{cc}}=R_\bb(1-x)$, we obtain
\begin{equation}
    \alpha \, \frac{1-2C_\bb}{C_\bb}
    \approx
    \ln\left(\frac{1}{1-x}\right).
\end{equation}
Solving for $x$, yields,
\begin{equation}
    \frac{\Delta R}{R_\bb}
    \approx
    1-\exp\!\left[
        - \alpha \, \frac{1-2C_\bb}{C_\bb}
    \right].
\label{eq:ct_integral}
\end{equation}
We call this is the ``exponential form.''

For a thin crust, $x\ll 1$, one may further use
\begin{equation}
    \ln\left(\frac{1}{1-x}\right) = -\ln(1-x) \approx x,
\end{equation}
which gives what we name the ``linearized form,''
\begin{equation}
\label{eq:crust_thickness}
    \frac{\Delta R}{R_\bb}
    \approx
    \alpha \, \frac{1-2C_\bb}{C_\bb}.
\end{equation}

Alternatively, using an approximation that we found yields better agreement with the numerical results,
\begin{equation}
    \ln\left(\frac{1}{1-x}\right)\approx \frac{x}{1-x},
\end{equation}
we obtain Eq.~\eqref{eq:ct_b6} in the main text, which we will 
call the ``rational form'' here.

This expression is equivalent in form to Eq.~(B6) of Ref.~\cite{Samuelsson:2006tt}, while the linearized form obtained earlier, Eq.~\eqref{eq:crust_thickness}, has the same form as Eq.~(B9) of Ref.~\cite{Samuelsson:2006tt}.

These expressions have the same form as those of Ref.~\cite{Samuelsson:2006tt}, up to the definition of the coefficient $\alpha$. We now show that, under the same polytropic crust assumptions they used, their parameter, which we denote by $\alpha_{\rm SA}$, is equivalent to our chemical-potential factor,
\begin{equation}
    \alpha_{\rm SA}
    =
    \ln({\mu_{\rm cc}}/{\mu_\su})
    =
    \alpha .
\end{equation}

To do so, we specialize to the same polytropic crust EOS used in Ref.~\cite{Samuelsson:2006tt}, Appendix B, 
\begin{equation}
    \varepsilon_\bb = k p_\bb^{1/\Gamma},
\end{equation}
where $k$ and $\Gamma$ are constants.
In the crust, where $p_\bb \ll \varepsilon_\bb$, the zero-temperature relation Eq.~\eqref{eq:zero_tem} 
gives
\begin{equation}
    \dd \ln\mu = \frac{\dd p_\bb}{\varepsilon_\bb+p_\bb} \approx \frac{\dd p_\bb}{\varepsilon_\bb(p_
    \bb)}.
\end{equation}
Integrating from the surface, where \(p_\bb=0\), to the crust-core interface, where \(p_\bb=p_{\rm cc}\), yields
\begin{equation}
    \alpha
    \approx
    \int^{p_{\bb,\rm cc}}_0\frac{\dd p_\bb}{\varepsilon_\bb(p_\bb)}.
\end{equation}
Substituting the polytropic equation of state, we obtain:
\begin{equation}
    \alpha
    \approx
    \int_0^{p_{\bb,\rm cc}}\frac{\dd p_\bb}{k p_\bb^{1/\Gamma}}
    =
    \frac{\Gamma}{\Gamma-1}\frac{1}{k} \, p_{\textrm{cc}}^{(\Gamma-1)/\Gamma}.
\end{equation}
Defining
\begin{equation}
    \chi = {\Gamma} / (\Gamma-1),
\end{equation}
so that $\chi^{-1}=(\Gamma-1)/\Gamma$, we obtain
\begin{equation}
\label{eq:poly_ref}
    \alpha
    =
    (\chi / k) \, p_{\bb,\rm cc}^{1/\chi}.
\end{equation}
Next, using the polytropic relation at the crust-core interface, we find
\begin{equation}
\label{eq:p_eps_ratio}
    \frac{p_{\bb,\textrm{cc}}}{\varepsilon_{\bb,\textrm{cc}}}
    =
    \frac{p_{\bb,\textrm{cc}}}{k p_{\bb,\textrm{cc}}^{1/\Gamma}}
    =
    \frac{1}{k} \, p_{\bb,\textrm{cc}}^{(\Gamma-1)/\Gamma}
    =
    \frac{1}{k} \, p_{\bb,\textrm{cc}}^{1/\chi}.
\end{equation}
Reference~\cite{Samuelsson:2006tt} defines
\begin{equation}
\label{eq:alphasa}
    \alpha_{\rm SA} = \chi \, {p_{\bb,\textrm{cc}}} / {\varepsilon_{\bb,\textrm{cc}}},
\end{equation}
where we have written their expression using our notation for the energy density, $\varepsilon_\bb$. Combining 
Eq.~\eqref{eq:alphasa} and~\eqref{eq:p_eps_ratio} we find that $\alpha =\alpha_{\rm SA}$,
as we wanted to show.

We remark that for the \eos equation of state, $\alpha = 0.0256$, which is close to the value $\alpha_{\rm SA} = 0.02326$ found by fitting over many equations of state in Ref.~\cite{Samuelsson:2006tt}.

\section{Numerical crust thickness} \label{app:ct_table}

In this Appendix, we compare the approximate analytical formulas for the crust thickness presented in Appendix~\ref{app:crust_thickness}, namely, the rational~\eqref{eq:ct_b6}, exponential~\eqref{eq:ct_integral}, and linearized~\eqref{eq:crust_thickness} expressions, against 
numerical data.

In Fig.~\ref{fig:ct_relerr}, we show the relative errors of the approximate crust thickness estimates with respect to the numerical results. Overall, the green dotted line, which is obtained with the linearized crust thickness estimate, Eq.~\eqref{eq:crust_thickness},  captures the qualitative trend that the fractional crust thickness decreases as $\mc$ increases. However, it systematically overestimates the numerical result. For $M_\tl=1.4\,\msun$, the relative error remains at the level of approximately $11$--$12\%$ over the full range of $m_\chi$, while for $M_\tl=2.0\,\msun$ the error is reduced to about $6$--$8\%;$.
The integral form approximation~\eqref{eq:ct_integral}, shown by the orange curve, gives better agreement with the numerical results, with relative errors of about $5\%$ for $M_\tl=1.4\,\msun$ and $2$--$4\%$ for $M_\tl=2.0\,\msun$.
The rational form approximation, ~\eqref{eq:ct_b6}, denoted by the blue curve, gives the best agreement among the three estimates. For $M_\tl=1.4\,\msun$, the relative error is typically below $0.5\%$, with a small systematic underestimation of the numerical result. For $M_\tl=2.0\,\msun$, the agreement is similarly good, with errors mostly below $0.5\%$ and only reaching about $1\%$ near $m_\chi=700\,{\rm MeV}$.

\section{Numerical torsional-mode frequencies} \label{app:tor_num_results}

\begin{figure}[t]
    \includegraphics[width=\columnwidth]{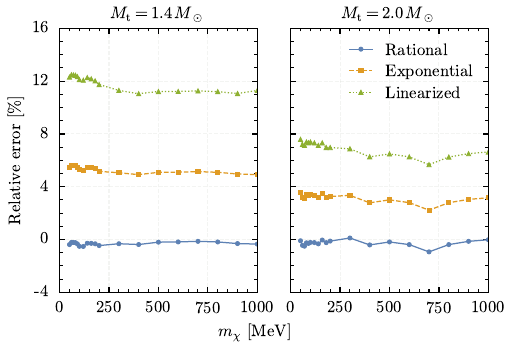}
    \caption{The relative error between the numerical crust thickness and the approximate estimates obtained from the rational~\eqref{eq:ct_b6}, exponential~\eqref{eq:ct_integral}, and linearized~\eqref{eq:crust_thickness} forms. 
    The left and right panels show the results for $M_\tl=1.4\,\msun$ and $M_\tl=2.0\,\msun$, respectively.
    The errors associated to each expression are indicated in the legend in the right panel. We see that in both panels, the rational form~\eqref{eq:ct_b6} approximated the best our numerical data.}
    \label{fig:ct_relerr}
\end{figure}

In this Appendix, we summarize the numerical torsional-mode frequencies  we obtained in Secs.~\ref{sec:results}
and~\ref{sec:results_es}. Table~\ref{tab:freq_nes_es_mt14_mt2} lists the fundamental mode ${}_{2}t_{0}$ and first overtone ${}_{2}t_{1}$ for \DANS{} models with $M_\tl=1.4\,\msun$ and $M_\tl=2.0\,\msun$, both at fixed 
dark matter fraction $f_\chi = 0.05$, vanishing self-interaction, and for a variety of $\mc$ values.
For each mode, we report the frequencies computed without electron screening (``NES'') and with electron screening (``ES'') included in the crust shear modulus.

\begingroup
\allowdisplaybreaks
\begin{table*}[t]
\centering
\caption{
Torsional oscillation frequencies ${}_{\ell} t_{n}$ for \DANS{} models constructed with the \eos baryonic equation of state and a fermionic dark matter equation of state with vanishing self-interaction, at fixed dark matter fraction $f_\chi = 0.05$.
For each entry, the value outside parentheses corresponds to $M_\tl=1.4\,\msun$, while the value in parentheses corresponds to $M_\tl=2.0\,\msun$.
The columns ${}_{2}t_{0}^{\mathrm{NES}}$ and ${}_{2}t_{0}^{\mathrm{ES}}$ denote the fundamental mode without and with electron screening, respectively, while ${}_{2}t_{1}^{\mathrm{NES}}$ and ${}_{2}t_{1}^{\mathrm{ES}}$ denote the corresponding first overtone.
}
\label{tab:freq_nes_es_mt14_mt2}
\begin{ruledtabular}
\begin{tabular}{ccccc}
$m_\chi$ [MeV]
& ${}_{2}t_{0}^{\mathrm{NES}}$ [Hz]
& ${}_{2}t_{0}^{\mathrm{ES}}$ [Hz]
& ${}_{2}t_{1}^{\mathrm{NES}}$ [Hz]
& ${}_{2}t_{1}^{\mathrm{ES}}$ [Hz] \\
\hline
50   & 33.318 (34.366) & 31.076 (32.048) & 672.971 (1034.052) & 632.035 (971.181) \\
60   & 33.318 (34.371) & 31.076 (32.055) & 673.133 (1034.483) & 632.186 (971.610) \\
70   & 33.321 (34.383) & 31.079 (32.065) & 673.374 (1035.700) & 632.413 (972.731) \\
80   & 33.327 (34.398) & 31.084 (32.080) & 673.764 (1037.219) & 632.778 (974.165) \\
90   & 33.332 (34.431) & 31.089 (32.110) & 674.268 (1039.563) & 633.253 (976.357) \\
100  & 33.341 (34.459) & 31.097 (32.136) & 675.013 (1042.690) & 633.951 (979.299) \\
120  & 33.369 (34.559) & 31.122 (32.231) & 677.236 (1052.274) & 636.041 (988.297) \\
140  & 33.407 (34.724) & 31.159 (32.383) & 680.750 (1067.100) & 639.341 (1002.225) \\
160  & 33.465 (34.945) & 31.213 (32.589) & 685.739 (1087.391) & 644.027 (1021.280) \\
180  & 33.540 (35.208) & 31.283 (32.833) & 692.234 (1111.004) & 650.127 (1043.468) \\
200  & 33.634 (35.482) & 31.370 (33.091) & 699.949 (1134.997) & 657.374 (1066.000) \\
300  & 34.130 (36.466) & 31.833 (34.006) & 738.700 (1210.033) & 693.772 (1136.471) \\
400  & 34.470 (37.005) & 32.150 (34.508) & 757.408 (1235.699) & 711.342 (1160.575) \\
500  & 34.699 (37.398) & 32.364 (34.876) & 764.305 (1251.334) & 717.817 (1175.263) \\
600  & 34.883 (37.714) & 32.535 (35.170) & 769.203 (1263.684) & 722.419 (1186.864) \\
700  & 35.043 (37.997) & 32.684 (35.436) & 773.280 (1274.386) & 726.249 (1196.916) \\
800  & 35.190 (38.265) & 32.821 (35.686) & 776.900 (1284.056) & 729.650 (1206.005) \\
900  & 35.331 (38.527) & 32.953 (35.930) & 780.299 (1293.471) & 732.841 (1214.839) \\
1000 & 35.471 (38.784) & 33.082 (36.170) & 783.551 (1302.672) & 735.896 (1223.479) \\
\end{tabular}
\end{ruledtabular}
\end{table*}
\endgroup

\clearpage

\bibliography{biblio}

@article{Kodama:PTP1972,
    author = {Kodama, Takeshi and Yamada, Masami},
    title = {Theory of Superdense Stars},
    journal = {Progress of Theoretical Physics},
    volume = {47},
    number = {2},
    pages = {444-459},
    year = {1972},
    month = {02},
    issn = {0033-068X},
    doi = {10.1143/PTP.47.444},
    url = {https://doi.org/10.1143/PTP.47.444},
}

@article{Narain:2006kx,
    author = "Narain, Gaurav and Schaffner-Bielich, Jurgen and Mishustin, Igor N.",
    title = "{Compact stars made of fermionic dark matter}",
    eprint = "astro-ph/0605724",
    archivePrefix = "arXiv",
    doi = "10.1103/PhysRevD.74.063003",
    journal = "Phys. Rev. D",
    volume = "74",
    pages = "063003",
    year = "2006"
}

@article{Pais:2016xiu,
    author = "Pais, Helena and Provid{\^e}ncia, Constan{\c{c}}a",
    title = "{Vlasov formalism for extended relativistic mean field models: The crust-core transition and the stellar matter equation of state}",
    eprint = "1607.05899",
    archivePrefix = "arXiv",
    primaryClass = "nucl-th",
    doi = "10.1103/PhysRevC.94.015808",
    journal = "Phys. Rev. C",
    volume = "94",
    number = "1",
    pages = "015808",
    year = "2016"
}

@article{Horowitz:2000xj,
    author = "Horowitz, C. J. and Piekarewicz, J.",
    title = "{Neutron star structure and the neutron radius of Pb-208}",
    eprint = "astro-ph/0010227",
    archivePrefix = "arXiv",
    doi = "10.1103/PhysRevLett.86.5647",
    journal = "Phys. Rev. Lett.",
    volume = "86",
    pages = "5647",
    year = "2001"
}

@article{Shawqi:2024jmk,
    author = "Shawqi, Shafayat and Morsink, Sharon M.",
    title = "{Interpreting Mass and Radius Measurements of Neutron Stars with Dark Matter Halos}",
    eprint = "2406.03332",
    archivePrefix = "arXiv",
    primaryClass = "astro-ph.HE",
    reportNumber = "INT-PUB-24-025",
    doi = "10.3847/1538-4357/ad77c1",
    journal = "Astrophys. J.",
    volume = "975",
    number = "1",
    pages = "123",
    year = "2024"
}

@article{Typel:2013rza,
    author = {Typel, S. and Oertel, M. and Kl{\"a}hn, T.},
    title = "{CompOSE CompStar online supernova equations of state harmonising the concert of nuclear physics and astrophysics compose.obspm.fr}",
    eprint = "1307.5715",
    archivePrefix = "arXiv",
    primaryClass = "astro-ph.SR",
    doi = "10.1134/S1063779615040061",
    journal = "Phys. Part. Nucl.",
    volume = "46",
    number = "4",
    pages = "633--664",
    year = "2015"
}

@article{Oertel:2016bki,
    author = {Oertel, M. and Hempel, M. and Kl{\"a}hn, T. and Typel, S.},
    title = "{Equations of state for supernovae and compact stars}",
    eprint = "1610.03361",
    archivePrefix = "arXiv",
    primaryClass = "astro-ph.HE",
    doi = "10.1103/RevModPhys.89.015007",
    journal = "Rev. Mod. Phys.",
    volume = "89",
    number = "1",
    pages = "015007",
    year = "2017"
}

@article{CompOSECoreTeam:2022ddl,
    author = "Typel, S. and others",
    collaboration = "CompOSE Core Team",
    title = "{CompOSE Reference Manual}",
    eprint = "2203.03209",
    archivePrefix = "arXiv",
    primaryClass = "astro-ph.HE",
    doi = "10.1140/epja/s10050-022-00847-y",
    journal = "Eur. Phys. J. A",
    volume = "58",
    number = "11",
    pages = "221",
    year = "2022"
}

@article{Samuelsson:2006tt,
    author = "Samuelsson, Lars and Andersson, Nils",
    title = "{Neutron Star Asteroseismology. Axial Crust Oscillations in the Cowling Approximation}",
    eprint = "astro-ph/0609265",
    archivePrefix = "arXiv",
    doi = "10.1111/j.1365-2966.2006.11147.x",
    journal = "Mon. Not. Roy. Astron. Soc.",
    volume = "374",
    pages = "256--268",
    year = "2007"
}

@article{Sotani:2014dua,
    author = "Sotani, Hajime",
    title = "{Electron screening effects on crustal torsional oscillations}",
    eprint = "1401.6977",
    archivePrefix = "arXiv",
    primaryClass = "astro-ph.HE",
    reportNumber = "YITP-14-8",
    doi = "10.1016/j.physletb.2014.01.054",
    journal = "Phys. Lett. B",
    volume = "730",
    pages = "166--170",
    year = "2014"
}

@article{Strohmayer:1991ApJ,
       author = {{Strohmayer}, T. and {Ogata}, S. and {Iyetomi}, H. and {Ichimaru}, S. and {van Horn}, H.~M.},
        title = "{The Shear Modulus of the Neutron Star Crust and Nonradial Oscillations of Neutron Stars}",
      journal = {\apj},
         year = 1991,
        month = jul,
       volume = {375},
        pages = {679},
          doi = {10.1086/170231},
}

@article{Sotani:2024mlb,
    author = "Sotani, Hajime",
    title = "{Magnetar QPOs and Neutron Star Crust Elasticity}",
    eprint = "2405.11858",
    archivePrefix = "arXiv",
    primaryClass = "astro-ph.HE",
    reportNumber = "RIKEN-iTHEMS-Report-24",
    doi = "10.3390/universe10060231",
    journal = "Universe",
    volume = "10",
    number = "6",
    pages = "231",
    year = "2024"
}

@article{Sandin:2008db,
    author = "Sandin, Fredrik and Ciarcelluti, Paolo",
    title = "{Effects of mirror dark matter on neutron stars}",
    eprint = "0809.2942",
    archivePrefix = "arXiv",
    primaryClass = "astro-ph",
    doi = "10.1016/j.astropartphys.2009.09.005",
    journal = "Astropart. Phys.",
    volume = "32",
    pages = "278--284",
    year = "2009"
}

@article{Ciarcelluti:2010ji,
    author = "Ciarcelluti, Paolo and Sandin, Fredrik",
    title = "{Have neutron stars a dark matter core?}",
    eprint = "1005.0857",
    archivePrefix = "arXiv",
    primaryClass = "astro-ph.HE",
    doi = "10.1016/j.physletb.2010.11.021",
    journal = "Phys. Lett. B",
    volume = "695",
    pages = "19--21",
    year = "2011"
}

@article{Kobyakov:2013eta,
    author = "Kobyakov, D. and Pethick, C. J.",
    title = "{Dynamics of the inner crust of neutron stars: hydrodynamics, elasticity and collective modes}",
    eprint = "1303.1315",
    archivePrefix = "arXiv",
    primaryClass = "nucl-th",
    doi = "10.1103/PhysRevC.87.055803",
    journal = "Phys. Rev. C",
    volume = "87",
    number = "5",
    pages = "055803",
    year = "2013",
    note = "[Erratum: Phys.Rev.C 94, 059902 (2016)]"
}

@article{Douchin:2001sv,
    author = "Douchin, F. and Haensel, P.",
    title = "{A unified equation of state of dense matter and neutron star structure}",
    eprint = "astro-ph/0111092",
    archivePrefix = "arXiv",
    doi = "10.1051/0004-6361:20011402",
    journal = "Astron. Astrophys.",
    volume = "380",
    pages = "151",
    year = "2001"
}

@article{Ogata:1990PhRvA,
       author = {{Ogata}, Shuji and {Ichimaru}, Setsuo},
        title = "{First-principles calculations of shear moduli for Monte Carlo-simulated Coulomb solids}",
      journal = {\pra},
         year = 1990,
        month = oct,
       volume = {42},
       number = {8},
        pages = {4867-4870},
          doi = {10.1103/PhysRevA.42.4867},
}

@article{Nelson:2018xtr,
    author = "Nelson, Ann and Reddy, Sanjay and Zhou, Dake",
    title = "{Dark halos around neutron stars and gravitational waves}",
    eprint = "1803.03266",
    archivePrefix = "arXiv",
    primaryClass = "hep-ph",
    reportNumber = "INT-PUB-18-010",
    doi = "10.1088/1475-7516/2019/07/012",
    journal = "JCAP",
    volume = "07",
    pages = "012",
    year = "2019"
}

@article{Tolman:1939jz,
    author = "Tolman, Richard C.",
    title = "{Static solutions of Einstein's field equations for spheres of fluid}",
    doi = "10.1103/PhysRev.55.364",
    journal = "Phys. Rev.",
    volume = "55",
    pages = "364--373",
    year = "1939"
}

@article{McDermott:1983ApJ,
       author = {{McDermott}, P.~N. and {van Horn}, H.~M. and {Scholl}, J.~F.},
        title = "{Nonradial g-mode oscillations of warm neutron stars}",
      journal = {\apj},
         year = 1983,
        month = may,
       volume = {268},
        pages = {837-848},
          doi = {10.1086/161006},
       adsurl = {https://ui.adsabs.harvard.edu/abs/1983ApJ...268..837M}
}

@misc{Watts:2011kh,
    author = "Watts, Anna L.",
    title = "{Neutron starquakes and the dynamic crust}",
    eprint = "1111.0514",
    archivePrefix = "arXiv",
    primaryClass = "astro-ph.SR",
    month = "11",
    year = "2011"
}

@article{Rutherford:2022xeb,
    author = "Rutherford, Nathan and Raaijmakers, Geert and Prescod-Weinstein, Chanda and Watts, Anna",
    title = "{Constraining bosonic asymmetric dark matter with neutron star mass-radius measurements}",
    eprint = "2208.03282",
    archivePrefix = "arXiv",
    primaryClass = "astro-ph.HE",
    doi = "10.1103/PhysRevD.107.103051",
    journal = "Phys. Rev. D",
    volume = "107",
    number = "10",
    pages = "103051",
    year = "2023"
}

@article{Rutherford:2024uix,
    author = "Rutherford, Nathan and Prescod-Weinstein, Chanda and Watts, Anna",
    title = "{Probing fermionic asymmetric dark matter cores using global neutron star properties}",
    eprint = "2410.00140",
    archivePrefix = "arXiv",
    primaryClass = "astro-ph.HE",
    doi = "10.1103/PhysRevD.111.123034",
    journal = "Phys. Rev. D",
    volume = "111",
    number = "12",
    pages = "123034",
    year = "2025"
}

@article{Kumar:2024zzl,
    author = "Kumar, Ankit and Sotani, Hajime",
    title = "{Constraints on the parameter space in dark matter admixed neutron stars}",
    eprint = "2408.15312",
    archivePrefix = "arXiv",
    primaryClass = "astro-ph.HE",
    reportNumber = "RIKEN-iTHEMS-Report-24",
    doi = "10.1103/PhysRevD.110.063001",
    journal = "Phys. Rev. D",
    volume = "110",
    number = "6",
    pages = "063001",
    year = "2024"
}

@article{Kozhberov:2020lzm,
    author = "Kozhberov, A. A. and Yakovlev, D. G.",
    title = "{Deformed crystals and torsional oscillations of neutron star crust}",
    eprint = "2009.04952",
    archivePrefix = "arXiv",
    primaryClass = "astro-ph.HE",
    doi = "10.1093/mnras/staa2715",
    journal = "Mon. Not. Roy. Astron. Soc.",
    volume = "498",
    number = "4",
    pages = "5149--5158",
    year = "2020"
}

@article{Yakovlev:2022nui,
    author = "Yakovlev, D. G.",
    title = "{Self-similarity relations for torsional oscillations of neutron stars}",
    eprint = "2210.02931",
    archivePrefix = "arXiv",
    primaryClass = "astro-ph.SR",
    doi = "10.1093/mnras/stac2871",
    journal = "Mon. Not. Roy. Astron. Soc.",
    volume = "518",
    number = "1",
    pages = "1148--1157",
    year = "2022"
}

@article{Glampedakis:2006apa,
    author = "Glampedakis, Kostas and Samuelsson, Lars and Andersson, Nils",
    title = "{Elastic or magnetic? A toy model for global magnetar oscillations with implications for QPOs during flares}",
    eprint = "astro-ph/0605461",
    archivePrefix = "arXiv",
    doi = "10.1111/j.1745-3933.2006.00211.x",
    journal = "Mon. Not. Roy. Astron. Soc.",
    volume = "371",
    pages = "L74--L77",
    year = "2006"
}

@article{Sotani:2024hsm,
    author = "Sotani, Hajime and Suvorov, Arthur G. and Kokkotas, Kostas D.",
    title = "{Couplings of torsional and shear oscillations in a neutron star crust}",
    eprint = "2406.17195",
    archivePrefix = "arXiv",
    primaryClass = "astro-ph.HE",
    reportNumber = "RIKEN-iTHEMS-Report-24",
    doi = "10.1103/PhysRevD.110.023035",
    journal = "Phys. Rev. D",
    volume = "110",
    number = "2",
    pages = "023035",
    year = "2024"
}

@article{Vavoulidis:2007ui,
    author = "Vavoulidis, M. and Stavridis, A. and Kokkotas, K. D. and Beyer, Horst R.",
    title = "{Torsional Oscillations of Slowly Rotating Relativistic Stars}",
    eprint = "gr-qc/0703039",
    archivePrefix = "arXiv",
    doi = "10.1111/j.1365-2966.2007.11706.x",
    journal = "Mon. Not. Roy. Astron. Soc.",
    volume = "377",
    pages = "1553--1556",
    year = "2007"
}

@article{Vavoulidis:2007cs,
    author = "Vavoulidis, M. and Kokkotas, K. D. and Stavridis, A.",
    title = "{Crustal Oscillations of Slowly Rotating Relativistic Stars}",
    eprint = "0712.1263",
    archivePrefix = "arXiv",
    primaryClass = "gr-qc",
    doi = "10.1111/j.1365-2966.2007.12835.x",
    journal = "Mon. Not. Roy. Astron. Soc.",
    volume = "384",
    pages = "1711",
    year = "2008"
}

@article{Liu:2024swd,
    author = "Liu, Yukun and Li, Hong-Bo and Gao, Yong and Shao, Lijing and Hu, Zexin",
    title = "{Effects from dark matter halos on x-ray pulsar pulse profiles}",
    eprint = "2408.04425",
    archivePrefix = "arXiv",
    primaryClass = "astro-ph.HE",
    doi = "10.1103/PhysRevD.110.083018",
    journal = "Phys. Rev. D",
    volume = "110",
    number = "8",
    pages = "083018",
    year = "2024"
}

@article{Finn:1988MNRAS,
    author = {Finn, Lee Samuel},
    title = "{Relativistic stellar pulsations in the Cowling approximation}",
    journal = {Monthly Notices of the Royal Astronomical Society},
    volume = {232},
    number = {2},
    pages = {259-275},
    year = {1988},
    month = {05},
    issn = {0035-8711},
    doi = {10.1093/mnras/232.2.259},
    url = {https://doi.org/10.1093/mnras/232.2.259},
}

@article{Burnaz:2025tza,
    author = "Burnaz, Louis and Most, Elias R. and Bransgrove, Ashley",
    title = "{Crustal Quakes Spark Magnetospheric Blasts: Imprints of Realistic Magnetar Crust Oscillations on the Fast Radio Burst Signal}",
    eprint = "2508.18033",
    archivePrefix = "arXiv",
    primaryClass = "astro-ph.HE",
    doi = "10.3847/2041-8213/ae2466",
    journal = "Astrophys. J. Lett.",
    volume = "995",
    number = "2",
    pages = "L57",
    year = "2025"
}

@article{Carter:1977qf,
    author = "Carter, B. and Quintana, H.",
    title = "{Gravitational and Acoustic Waves in an Elastic Medium}",
    doi = "10.1103/PhysRevD.16.2928",
    journal = "Phys. Rev. D",
    volume = "16",
    pages = "2928--2938",
    year = "1977"
}

@misc{thorne:1967ApJ,
       author = {{Thorne}, Kip S. and {Campolattaro}, Alfonso},
        title = "{Non-Radial Pulsation of General-Relativistic Stellar Models. I. Analytic Analysis for $l \geq 2$}",
         year = 1967,
        month = sep,
        pages = {591},
          doi = {10.1086/149288},
    publisher = {IOP},
       adsurl = {https://ui.adsabs.harvard.edu/abs/1967ApJ...149..591T}
}

@article{Oppenheimer:1939ne,
    author = "Oppenheimer, J. R. and Volkoff, G. M.",
    title = "{On massive neutron cores}",
    doi = "10.1103/PhysRev.55.374",
    journal = "Phys. Rev.",
    volume = "55",
    pages = "374--381",
    year = "1939"
}

@article{Ellis:2018bkr,
    author = {Ellis, John and H{\"u}tsi, Gert and Kannike, Kristjan and Marzola, Luca and Raidal, Martti and Vaskonen, Ville},
    title = "{Dark Matter Effects On Neutron Star Properties}",
    eprint = "1804.01418",
    archivePrefix = "arXiv",
    primaryClass = "astro-ph.CO",
    reportNumber = "CERN-TH-2018-072, KCL-PH-TH/2018-13, KCL-PH-TH-2018-13",
    doi = "10.1103/PhysRevD.97.123007",
    journal = "Phys. Rev. D",
    volume = "97",
    number = "12",
    pages = "123007",
    year = "2018"
}

@article{Schumaker:1983MNRAS.203..457S,
       author = {{Schumaker}, B.~L. and {Thorne}, K.~S.},
        title = "{Torsional oscillations of neutron stars}",
      journal = {MNRAS},
         year = 1983,
        month = may,
       volume = {203},
        pages = {457-489},
          doi = {10.1093/mnras/203.2.457},
       adsurl = {https://ui.adsabs.harvard.edu/abs/1983MNRAS.203..457S}
}

@article{Leung:2022wcf,
    author = "Leung, Kwing-Lam and Chu, Ming-chung and Lin, Lap-Ming",
    title = "{Tidal deformability of dark matter admixed neutron stars}",
    eprint = "2207.02433",
    archivePrefix = "arXiv",
    primaryClass = "astro-ph.HE",
    doi = "10.1103/PhysRevD.105.123010",
    journal = "Phys. Rev. D",
    volume = "105",
    number = "12",
    pages = "123010",
    year = "2022"
}

@article{Chamel:2008ca,
    author = "Chamel, N. and Haensel, P.",
    title = "{Physics of Neutron Star Crusts}",
    eprint = "0812.3955",
    archivePrefix = "arXiv",
    primaryClass = "astro-ph",
    doi = "10.12942/lrr-2008-10",
    journal = "Living Rev. Rel.",
    volume = "11",
    pages = "10",
    year = "2008"
}

@article{Fortin:2016hny,
    author = "Fortin, M. and Providencia, C. and Raduta, A. R. and Gulminelli, F. and Zdunik, J. L and Haensel, P. and Bejger, M.",
    title = "{Neutron star radii and crusts: uncertainties and unified equations of state}",
    eprint = "1604.01944",
    archivePrefix = "arXiv",
    primaryClass = "astro-ph.SR",
    doi = "10.1103/PhysRevC.94.035804",
    journal = "Phys. Rev. C",
    volume = "94",
    number = "3",
    pages = "035804",
    year = "2016"
}

@article{Baym:2017whm,
    author = "Baym, Gordon and Hatsuda, Tetsuo and Kojo, Toru and Powell, Philip D. and Song, Yifan and Takatsuka, Tatsuyuki",
    title = "{From hadrons to quarks in neutron stars: a review}",
    eprint = "1707.04966",
    archivePrefix = "arXiv",
    primaryClass = "astro-ph.HE",
    reportNumber = "RIKEN-ITHEMS-REPORT-17, RIKEN-QHP-316, RIKEN-iTHEMS-Report-17",
    doi = "10.1088/1361-6633/aaae14",
    journal = "Rept. Prog. Phys.",
    volume = "81",
    number = "5",
    pages = "056902",
    year = "2018"
}

@article{Henriques:1990xg,
    author = "Henriques, A. B. and Liddle, Andrew R. and Moorhouse, R. G.",
    title = "{Stability of boson-fermion stars}",
    doi = "10.1016/0370-2693(90)90789-9",
    journal = "Phys. Lett. B",
    volume = "251",
    pages = "511--516",
    year = "1990"
}

@article{Jetzer:1990xa,
    author = "Jetzer, P.",
    title = "{Stability of combined boson-fermion stars}",
    reportNumber = "CERN-TH-5685-90",
    doi = "10.1016/0370-2693(90)90952-3",
    journal = "Phys. Lett. B",
    volume = "243",
    pages = "36--40",
    year = "1990"
}

@article{Sorkin:1982ut,
    author = "Sorkin, Rafael D.",
    title = "{A Stability criterion for many parameter equilibrium families}",
    reportNumber = "PRINT-81-0901 (IAS,PRINCETON)",
    doi = "10.1086/160034",
    journal = "Astrophys. J.",
    volume = "257",
    pages = "847--854",
    year = "1982"
}

@article{Caballero:2024qtv,
    author = "Caballero, Daniel A. and Ripley, Justin L. and Yunes, Nicol{\'a}s",
    title = "{Radial mode stability of two-fluid neutron stars}",
    eprint = "2408.04701",
    archivePrefix = "arXiv",
    primaryClass = "gr-qc",
    doi = "10.1103/PhysRevD.110.103038",
    journal = "Phys. Rev. D",
    volume = "110",
    number = "10",
    pages = "103038",
    year = "2024"
}

@article{Kumar:2025ytm,
    author = "Kumar, Ankit and Sotani, Hajime",
    title = "{Impact of dark matter distribution on neutron star properties}",
    eprint = "2501.07052",
    archivePrefix = "arXiv",
    primaryClass = "astro-ph.HE",
    reportNumber = "RIKEN-iTHEMS-Report-25",
    doi = "10.1103/PhysRevD.111.043016",
    journal = "Phys. Rev. D",
    volume = "111",
    number = "4",
    pages = "043016",
    year = "2025"
}

@article{Goldman:1989nd,
    author = "Goldman, I. and Nussinov, S.",
    title = "{Weakly Interacting Massive Particles and Neutron Stars}",
    doi = "10.1103/PhysRevD.40.3221",
    journal = "Phys. Rev. D",
    volume = "40",
    pages = "3221--3230",
    year = "1989"
}

@article{Bell:2020jou,
    author = "Bell, Nicole F. and Busoni, Giorgio and Robles, Sandra and Virgato, Michael",
    title = "{Improved Treatment of Dark Matter Capture in Neutron Stars}",
    eprint = "2004.14888",
    archivePrefix = "arXiv",
    primaryClass = "hep-ph",
    doi = "10.1088/1475-7516/2020/09/028",
    journal = "JCAP",
    volume = "09",
    pages = "028",
    year = "2020"
}

@article{Busoni:2021zoe,
    author = "Busoni, Giorgio",
    title = "{Capture of Dark Matter in Neutron Stars}",
    eprint = "2201.00048",
    archivePrefix = "arXiv",
    primaryClass = "hep-ph",
    doi = "10.3103/S0027134922020205",
    journal = "Moscow Univ. Phys. Bull.",
    volume = "77",
    number = "2",
    pages = "301--305",
    year = "2022"
}

@article{Kain:2021hpk,
    author = "Kain, Ben",
    title = "{Dark matter admixed neutron stars}",
    eprint = "2102.08257",
    archivePrefix = "arXiv",
    primaryClass = "gr-qc",
    doi = "10.1103/PhysRevD.103.043009",
    journal = "Phys. Rev. D",
    volume = "103",
    number = "4",
    pages = "043009",
    year = "2021"
}

@article{Karkevandi:2021ygv,
    author = "Karkevandi, Davood Rafiei and Shakeri, Soroush and Sagun, Violetta and Ivanytskyi, Oleksii",
    title = "{Bosonic dark matter in neutron stars and its effect on gravitational wave signal}",
    eprint = "2109.03801",
    archivePrefix = "arXiv",
    primaryClass = "astro-ph.HE",
    doi = "10.1103/PhysRevD.105.023001",
    journal = "Phys. Rev. D",
    volume = "105",
    number = "2",
    pages = "023001",
    year = "2022"
}

@article{Leung:2011zz,
    author = "Leung, S. C. and Chu, M. C. and Lin, L. M.",
    title = "{Dark-matter admixed neutron stars}",
    eprint = "1111.1787",
    archivePrefix = "arXiv",
    primaryClass = "astro-ph.CO",
    doi = "10.1103/PhysRevD.84.107301",
    journal = "Phys. Rev. D",
    volume = "84",
    pages = "107301",
    year = "2011"
}

@article{Das:2020ecp,
    author = "Das, Arpan and Malik, Tuhin and Nayak, Alekha C.",
    title = "{Dark matter admixed neutron star properties in light of gravitational wave observations: A two fluid approach}",
    eprint = "2011.01318",
    archivePrefix = "arXiv",
    primaryClass = "nucl-th",
    doi = "10.1103/PhysRevD.105.123034",
    journal = "Phys. Rev. D",
    volume = "105",
    number = "12",
    pages = "123034",
    year = "2022"
}

@article{Karkevandi:2024vov,
    author = "Karkevandi, Davood Rafiei and Shahrbaf, Mahboubeh and Shakeri, Soroush and Typel, Stefan",
    title = "{Exploring the Distribution and Impact of Bosonic Dark Matter in Neutron Stars}",
    eprint = "2402.18696",
    archivePrefix = "arXiv",
    primaryClass = "astro-ph.HE",
    doi = "10.3390/particles7010011",
    journal = "Particles",
    volume = "7",
    number = "1",
    pages = "201--213",
    year = "2024"
}

@article{Miao:2022rqj,
    author = "Miao, Zhiqiang and Zhu, Yaofeng and Li, Ang and Huang, Feng",
    title = "{Dark Matter Admixed Neutron Star Properties in the Light of X-Ray Pulse Profile Observations}",
    eprint = "2204.05560",
    archivePrefix = "arXiv",
    primaryClass = "astro-ph.HE",
    doi = "10.3847/1538-4357/ac8544",
    journal = "Astrophys. J.",
    volume = "936",
    number = "1",
    pages = "69",
    year = "2022"
}

@article{Sagun:2022ezx,
    author = "Sagun, Violetta and Giangrandi, Edoardo and Ivanytskyi, Oleksii and Provid{\^e}ncia, Constan{\c{c}}a and Dietrich, Tim",
    title = "{How does dark matter affect compact star properties and high density constraints of strongly interacting matter}",
    eprint = "2211.10510",
    archivePrefix = "arXiv",
    primaryClass = "astro-ph.HE",
    doi = "10.1051/epjconf/202227407009",
    journal = "EPJ Web Conf.",
    volume = "274",
    pages = "07009",
    year = "2022"
}

@article{Routaray:2022utr,
    author = "Routaray, Pinku and Das, H. C. and Sen, Souhardya and Kumar, Bharat and Panotopoulos, Grigoris and Zhao, Tianqi",
    title = "{Radial oscillations of dark matter admixed neutron stars}",
    eprint = "2211.12808",
    archivePrefix = "arXiv",
    primaryClass = "nucl-th",
    doi = "10.1103/PhysRevD.107.103039",
    journal = "Phys. Rev. D",
    volume = "107",
    number = "10",
    pages = "103039",
    year = "2023"
}

@article{Leung:2012vea,
    author = "Leung, S. C. and Chu, M. C. and Lin, L. M.",
    title = "{Equilibrium Structure and Radial Oscillations of Dark Matter Admixed Neutron Stars}",
    eprint = "1205.1909",
    archivePrefix = "arXiv",
    primaryClass = "astro-ph.CO",
    doi = "10.1103/PhysRevD.85.103528",
    journal = "Phys. Rev. D",
    volume = "85",
    pages = "103528",
    year = "2012"
}

@article{Sotani:2025hzb,
    author = "Sotani, Hajime and Kumar, Ankit",
    title = "{Emergence of new oscillation modes in dark matter admixed neutron stars}",
    eprint = "2505.18800",
    archivePrefix = "arXiv",
    primaryClass = "astro-ph.HE",
    reportNumber = "RIKEN-iTHEMS-Report-25",
    doi = "10.1103/kcl2-qgxh",
    journal = "Phys. Rev. D",
    volume = "111",
    number = "12",
    pages = "123013",
    year = "2025"
}

@article{Strohmayer:2006py,
    author = "Strohmayer, Tod E. and Watts, Anna L.",
    title = "{The 2004 Hyperflare from SGR 1806-20: Further Evidence for Global Torsional Vibrations}",
    eprint = "astro-ph/0608463",
    archivePrefix = "arXiv",
    doi = "10.1086/508703",
    journal = "Astrophys. J.",
    volume = "653",
    pages = "593--601",
    year = "2006"
}

@article{Watts:2006mr,
    author = "Watts, Anna L. and Strohmayer, Tod E.",
    title = "{Neutron star oscillations and QPOs during magnetar flares}",
    eprint = "astro-ph/0612252",
    archivePrefix = "arXiv",
    doi = "10.1016/j.asr.2006.12.021",
    journal = "Adv. Space Res.",
    volume = "40",
    pages = "1446--1452",
    year = "2007"
}

@article{Sotani:2012qc,
    author = "Sotani, Hajime and Nakazato, Ken'ichiro and Iida, Kei and Oyamatsu, Kazuhiro",
    title = "{Probing the Equation of State of Nuclear Matter via Neutron Star Asteroseismology}",
    eprint = "1202.6242",
    archivePrefix = "arXiv",
    primaryClass = "astro-ph.HE",
    doi = "10.1103/PhysRevLett.108.201101",
    journal = "Phys. Rev. Lett.",
    volume = "108",
    pages = "201101",
    year = "2012"
}

@article{Gleason:2022eeg,
    author = "Gleason, Troy and Brown, Ben and Kain, Ben",
    title = "{Dynamical evolution of dark matter admixed neutron stars}",
    eprint = "2201.02274",
    archivePrefix = "arXiv",
    primaryClass = "gr-qc",
    doi = "10.1103/PhysRevD.105.023010",
    journal = "Phys. Rev. D",
    volume = "105",
    number = "2",
    pages = "023010",
    year = "2022"
}

@article{Potekhin:2013qqa,
    author = "Potekhin, A. Y. and Fantina, A. F. and Chamel, N. and Pearson, J. M. and Goriely, S.",
    title = "{Analytical representations of unified equations of state for neutron-star matter}",
    eprint = "1310.0049",
    archivePrefix = "arXiv",
    primaryClass = "astro-ph.SR",
    doi = "10.1051/0004-6361/201321697",
    journal = "Astron. Astrophys.",
    volume = "560",
    pages = "A48",
    year = "2013"
}

@article{Sotani:2018tdr,
    author = "Sotani, Hajime and Iida, Kei and Oyamatsu, Kazuhiro",
    title = "{Constraints on the Nuclear Equation of State and the Neutron Star Structure from Crustal Torsional Oscillations}",
    eprint = "1807.00528",
    archivePrefix = "arXiv",
    primaryClass = "astro-ph.HE",
    doi = "10.1093/mnras/sty1755",
    journal = "Mon. Not. Roy. Astron. Soc.",
    volume = "479",
    number = "4",
    pages = "4735--4748",
    year = "2018"
}

@article{Sotani:2006at,
    author = "Sotani, H. and Kokkotas, K. D. and Stergioulas, N.",
    title = "{Torsional Oscillations of Relativistic Stars with Dipole Magnetic Fields}",
    eprint = "astro-ph/0608626",
    archivePrefix = "arXiv",
    doi = "10.1111/j.1365-2966.2006.11304.x",
    journal = "Mon. Not. Roy. Astron. Soc.",
    volume = "375",
    pages = "261--277",
    year = "2007"
}

@article{deSouza:2018ope,
    author = "de Souza, Gibran H. and Chirenti, Cecilia",
    title = "{Torsional oscillations of magnetized neutron stars with mixed poloidal-toroidal fields}",
    eprint = "1810.06628",
    archivePrefix = "arXiv",
    primaryClass = "astro-ph.HE",
    doi = "10.1103/PhysRevD.100.043017",
    journal = "Phys. Rev. D",
    volume = "100",
    number = "4",
    pages = "043017",
    year = "2019"
}

@article{Messios:2001br,
    author = "Messios, Neophytos and Papadopoulos, Demetrios B. and Stergioulas, Nikolaos",
    title = "{Torsional oscillations of magnetized relativistic stars}",
    eprint = "astro-ph/0105175",
    archivePrefix = "arXiv",
    doi = "10.1046/j.1365-8711.2001.04645.x",
    journal = "Mon. Not. Roy. Astron. Soc.",
    volume = "328",
    pages = "1161",
    year = "2001"
}

@article{Kruger:2024fxn,
    author = {Kr{\"u}ger, Christian J. and Sotani, Hajime},
    title = "{Impact of the relativistic Cowling approximation on shear and interface modes of neutron stars}",
    eprint = "2411.03940",
    archivePrefix = "arXiv",
    primaryClass = "gr-qc",
    reportNumber = "RIKEN-iTHEMS-Report-24",
    doi = "10.1103/PhysRevD.111.063029",
    journal = "Phys. Rev. D",
    volume = "111",
    number = "6",
    pages = "063029",
    year = "2025"
}

@article{Yoshida:2002vd,
    author = "Yoshida, Shijun and Lee, Umin",
    title = "{Nonradial oscillations of neutron stars with a solid crust - analysis in the relativistic cowling approximation-}",
    eprint = "astro-ph/0210591",
    archivePrefix = "arXiv",
    doi = "10.1051/0004-6361:20021270",
    journal = "Astron. Astrophys.",
    volume = "395",
    pages = "201--208",
    year = "2002"
}

@article{1990MNRAS.245...82F,
       author = {{Finn}, Lee Samuel},
        title = "{Non-radial pulsations of neutron stars with a crust}",
      journal = {MNRAS},
         year = 1990,
        month = jul,
       volume = {245},
        pages = {82},
          doi = {10.1093/mnras/245.1.82},
       adsurl = {https://ui.adsabs.harvard.edu/abs/1990MNRAS.245...82F}
}

@article{El-Mezeini:2010xxh,
    author = "El-Mezeini, Ahmed M. and Ibrahim, Alaa I.",
    title = "{Discovery of Quasi-Periodic Oscillations in the Recurrent Burst Emission from SGR 1806-20}",
    eprint = "1008.3870",
    archivePrefix = "arXiv",
    primaryClass = "astro-ph.HE",
    doi = "10.1088/2041-8205/721/2/L121",
    journal = "Astrophys. J. Lett.",
    volume = "721",
    pages = "L121--L125",
    year = "2010"
}

@article{Israel:2005av,
    author = "Israel, GianLuca and Belloni, Tomaso and Stella, Luigi and Rephaeli, Yoel and Gruber, Duane and Casella, Pier Giorgio and Dall'Osso, Simone and Rea, Nanda and Persic, Massimo and Rothschild, Richard",
    title = "{Discovery of rapid x-ray oscillations in the tail of the SGR 1806-20 hyperflare}",
    eprint = "astro-ph/0505255",
    archivePrefix = "arXiv",
    doi = "10.1086/432615",
    journal = "Astrophys. J. Lett.",
    volume = "628",
    pages = "L53--L56",
    year = "2005"
}

@article{Duncan:1998my,
    author = "Duncan, Robert C.",
    title = "{Global seismic oscillations in soft gamma repeaters}",
    eprint = "astro-ph/9803060",
    archivePrefix = "arXiv",
    doi = "10.1086/311303",
    journal = "Astrophys. J. Lett.",
    volume = "498",
    pages = "L45",
    year = "1998"
}

@article{Levin:2006qd,
    author = "Levin, Yuri",
    title = "{On the theory of magnetar QPOs}",
    eprint = "astro-ph/0612725",
    archivePrefix = "arXiv",
    doi = "10.1111/j.1365-2966.2007.11582.x",
    journal = "Mon. Not. Roy. Astron. Soc.",
    volume = "377",
    pages = "159--167",
    year = "2007"
}

@article{Shawqi:2025cca,
    author = "Shawqi, Shafayat and Konstantinou, Andreas and Morsink, Sharon M.",
    title = "{Rotating neutron stars with dark matter halos}",
    eprint = "2508.18434",
    archivePrefix = "arXiv",
    primaryClass = "astro-ph.HE",
    doi = "10.1088/1475-7516/2026/04/011",
    journal = "JCAP",
    volume = "04",
    pages = "011",
    year = "2026"
}

@article{Lattimer:2004pg,
    author = "Lattimer, J. M. and Prakash, M.",
    title = "{The physics of neutron stars}",
    eprint = "astro-ph/0405262",
    archivePrefix = "arXiv",
    doi = "10.1126/science.1090720",
    journal = "Science",
    volume = "304",
    pages = "536--542",
    year = "2004"
}

@article{Ozel:2016oaf,
    author = {{\"O}zel, Feryal and Freire, Paulo},
    title = "{Masses, Radii, and the Equation of State of Neutron Stars}",
    eprint = "1603.02698",
    archivePrefix = "arXiv",
    primaryClass = "astro-ph.HE",
    doi = "10.1146/annurev-astro-081915-023322",
    journal = "Ann. Rev. Astron. Astrophys.",
    volume = "54",
    pages = "401--440",
    year = "2016"
}

@article{Lattimer:2021emm,
    author = "Lattimer, J. M.",
    title = "{Neutron Stars and the Nuclear Matter Equation of State}",
    doi = "10.1146/annurev-nucl-102419-124827",
    journal = "Ann. Rev. Nucl. Part. Sci.",
    volume = "71",
    pages = "433--464",
    year = "2021"
}

@article{Zhou:2025dmy,
    author = "Zhou, B. X. and Das, H. C. and Wei, J. B. and Burgio, G. F. and Li, Z. H. and Schulze, H. -J.",
    title = "{Cooling of dark neutron stars}",
    eprint = "2508.09704",
    archivePrefix = "arXiv",
    primaryClass = "astro-ph.HE",
    doi = "10.1103/nzq6-llbf",
    journal = "Phys. Rev. D",
    volume = "112",
    number = "12",
    pages = "123035",
    year = "2025"
}

@article{Routaray:2024fcq,
    author = "Routaray, Pinku and Parmar, Vishal and Das, H. C. and Kumar, Bharat and Burgio, G. F. and Schulze, H. -J.",
    title = "{Effects of asymmetric dark matter on a magnetized neutron star: A two-fluid approach}",
    eprint = "2412.21097",
    archivePrefix = "arXiv",
    primaryClass = "nucl-th",
    doi = "10.1103/PhysRevD.111.103045",
    journal = "Phys. Rev. D",
    volume = "111",
    number = "10",
    pages = "103045",
    year = "2025"
}

@article{Grippa:2024ach,
    author = "Grippa, Francesco and Lambiase, Gaetano and Poddar, Tanmay Kumar",
    title = "{Searching for New Physics in an Ultradense Environment: A Review on Dark Matter Admixed Neutron Stars}",
    eprint = "2412.09381",
    archivePrefix = "arXiv",
    primaryClass = "astro-ph.HE",
    doi = "10.3390/universe11030074",
    journal = "Universe",
    volume = "11",
    number = "3",
    pages = "74",
    year = "2025"
}

@article{Konstantinou:2024ynd,
    author = "Konstantinou, Andreas",
    title = "{The Effect of a Dark Matter Core on the Structure of a Rotating Neutron Star}",
    eprint = "2405.01487",
    archivePrefix = "arXiv",
    primaryClass = "astro-ph.HE",
    doi = "10.3847/1538-4357/ad4701",
    journal = "Astrophys. J.",
    volume = "968",
    number = "2",
    pages = "83",
    year = "2024"
}

\end{document}